\documentclass{article}

\usepackage[a4paper, total={6.5in, 8.5in}]{geometry}
\usepackage{graphicx}
\usepackage{authblk}
\usepackage{soul}
\usepackage{amsmath}
\usepackage[
backend=biber,
giveninits=true,
style=numeric-comp,
sorting=none,
maxcitenames=1,
natbib=true,
sortlocale=en_US,
url=false, 
doi=true,
eprint=false
]{biblatex}
\usepackage{geometry}
\usepackage{atbegshi}
\usepackage{float,lscape}
\usepackage{multirow}
\usepackage{blindtext}
\usepackage{amssymb}
\usepackage{ragged2e}
\usepackage{blindtext}
\usepackage{longtable}
\usepackage{comment}
\usepackage{xcolor}
\usepackage{caption}
\usepackage{bm}
\usepackage{changepage}

\usepackage{amsfonts}
\usepackage{tabularx}
\usepackage{lipsum}
\usepackage[title]{appendix}
\usepackage{mathtools}
\usepackage{circuitikz}
\usetikzlibrary{patterns}
\usepackage{hyperref}
\usepackage{caption}
\usepackage{booktabs}
\usepackage{array}
\usepackage{siunitx}
\usepackage{subcaption}
\usepackage{amsthm}
\newtheorem{theorem}{Theorem}

\hypersetup{
	colorlinks   = true, 
	urlcolor     = blue, 
	linkcolor    = blue, 
	citecolor   = red 
}
\providecommand{\keywords}[1]
{
  \small	
  \textbf{\textit{Keywords---}} #1
}

\title{Multi-Head Neural Operator for Modelling Interfacial Dynamics}

\author[1]{Mohammad Sadegh Eshaghi\thanks{Corresponding author. Email: eshaghi.khanghah@iop.uni-hannover.de}}
\author[1]{Navid Valizadeh}
\author[1]{Cosmin Anitescu}
\author[3]{Yizheng Wang}
\author[1]{Xiaoying Zhuang}
\author[2]{Timon Rabczuk\thanks{Corresponding author. Email: timon.rabczuk@uni-weimar.de}}

\affil[1]{Chair of Computational Science and Simulation Technology, Institute of Photonics, Department of Mathematics and Physics, Leibniz University Hannover, 30167 Hannover, Germany}
\affil[2]{Institute of Structural Mechanics, Bauhaus-Universität Weimar, Germany}
\affil[3]{Department of Engineering Mechanics, Tsinghua University, Beijing, China}

\date{}

\begin{document}

\maketitle

\begin{abstract}
Interfacial dynamics underlie a wide range of phenomena, including phase transitions, microstructure coarsening, pattern formation, and thin‑film growth, and are typically described by stiff, time-dependent nonlinear partial differential equations (PDEs). Traditional numerical methods, including finite difference, finite element, and spectral techniques, often become computationally prohibitive when dealing with high-dimensional problems or systems with multiple scales. Neural operators (NOs), a class of deep learning models, have emerged as a promising alternative by learning mappings between function spaces and efficiently approximating solution operators. In this work, we introduce the Multi-Head Neural Operator (MHNO), an extended neural operator framework specifically designed to address the temporal challenges associated with solving time-dependent PDEs. Unlike existing neural operators, which either struggle with error accumulation or require substantial computational resources for high-dimensional tensor representations, MHNO employs a novel architecture with time-step-specific projection operators and explicit temporal connections inspired by message-passing mechanisms. This design allows MHNO to predict all time steps after a single forward pass, while effectively capturing long-term dependencies and avoiding parameter overgrowth. We apply MHNO to solve various phase field equations, including antiphase boundary motion, spinodal decomposition, pattern formation, atomic scale modeling, and molecular beam epitaxy growth model, and compare its performance with existing NO-based methods. Our results show that MHNO achieves superior accuracy, scalability, and efficiency, demonstrating its potential as a next-generation computational tool for phase field modeling. The code and data supporting this work is publicly available at \href{https://github.com/eshaghi-ms/MHNO}{https://github.com/eshaghi-ms/MHNO}.
\end{abstract}


\section{Introduction} \label{sec:Introduction}

Phase field simulations are a cornerstone for modeling and understanding complex physical processes and interfacial dynamics in various scientific and engineering disciplines \cite{hashemi2021pore, moelans2008introduction, steinbach2009phase}. This method facilitates modeling complex interfacial dynamics\cite{zhu2020interfacial}, such as solidification \cite{kavousi2021quantitative}, fracture \cite{zhuang2022phase}, phase separation\cite{tegze2005phase}, vesicle dynamics \cite{valizadeh2022isogeometric, ashour2023phase, wen2024hydrodynamics, valizadeh2025monolithic}, and microstructure evolution in materials \cite{chen2002phase}, without requiring sophisticated algorithms for tracking interfaces \cite{tezduyar2006interface}. By introducing auxiliary scalar field variables, known as phase-field variables, to represent different phases or states, phase-field methods allow for the continuous description of sharp interfaces \cite{valizadeh2019isogeometric}.

Despite the advantages of the phase field method in modeling interfacial dynamics through a diffusive interface approach, its application is often limited by significant computational cost \cite{montes2021accelerating}. Traditional numerical techniques such as finite differences, spectral approximations \cite{xiao2022spectral}, mixed formulation finite element methods \cite{cervera2022comparative}, and isogeometric analysis \cite{valizadeh2022isogeometric, valizadeh2019isogeometric} are frequently used to solve the phase field equations, such as the fourth-order parabolic Cahn-Hilliard equation. However, these methods are restricted by the stiff nature of the equations, which must simultaneously resolve rapid phase separation and slow coalescence over different spatial and temporal scales \cite{oommen2022learning}. Achieving accurate results requires fine spatial and temporal discretizations, which significantly increases the computational cost. Computational cost becomes an even greater bottleneck in applications that require repeated simulations, such as uncertainty quantification \cite{smith2024uncertainty}, inverse problems \cite{galbally2010non}, and design optimization \cite{white2020dual}. These challenges have motivated the development of surrogate models, which are faster than high-fidelity simulations but often come at the expense of reduced accuracy \cite{bonneville2024accelerating}.

To circumvent the computational expense of phase-field simulations, strategies such as adaptive meshing \cite{george2002efficient} have been proposed. While effective in some scenarios, these methods have limitations when applied to simulations with complex and evolving morphologies. To overcome these limitations, machine learning-based methods have emerged as promising alternatives for accelerating phase field simulations \cite{kiyani2022machine}. These approaches fall broadly into two categories: data-driven finite-dimensional operators \cite{zhu2018bayesian, mo2019deep, zhong2019predicting, tang2020deep, wen2019multiphase, wen2021ccsnet, hu2022accelerating, montes2021accelerating, lee2020model, bonneville2024gplasdi, bonneville2024comprehensive, chakraborty2022domain}, which learn mappings between Euclidean spaces from numerical simulation data, and Physics-Informed versions \cite{raissi2019physics, zhu2019physics, haghighat2021sciann, chakraborty2022variational, goswami2020adaptive, goswami2020transfer, es4846935deepnetbeam}, which parameterize solution functions while embedding physical laws in their architectures. 

The first category, data-driven finite-dimensional operators, is often implemented using CNNs, which have shown great promise in providing fast and accurate predictions \cite{wen2019multiphase, jiang2021deep, wen2021ccsnet, tang2021deep, wu2021physics, geng2024deep}.  However, CNN-based approaches face several challenges. One major problem is their susceptibility to overfitting, especially when trained on limited datasets. As a result, they often require large numerical simulation datasets, which can become computationally prohibitive as the problem dimensionality increases \cite{oommen2024rethinking}. In addition, the predictions produced by these models are closely tied to the spatial and temporal discretizations of the training data, limiting their generalizability across different resolutions. This dependency on specific network configurations can lead to high generalization errors when the trained model is applied to unseen inputs with different discretizations or parameter settings. Consequently, while CNN-based finite-dimensional operators have proven useful in certain contexts, their reliance on extensive data and lack of resolution invariance present significant barriers to broader applicability \cite{wen2022u}. The second category, PINNs, incorporates physical laws into the network architecture to enforce solutions that satisfy governing equations such as PDEs. While PINNs have demonstrated success in many applications \cite{kamrava2021simulating, wang2021physics}, they face significant challenges in resolving sharp interfaces, which are common in phase field problems \cite{de2024physics, krishnapriyan2021characterizing, mao2023physics, you2022nonlocal}. Adaptive sampling methods \cite{wight2020solving} have been introduced to address this limitation by dynamically selecting collocation points in regions with steep gradients. However, such methods fail to generalize effectively to a variety of initial conditions and require retraining for each new initial condition, resulting in high computational costs for phase field models \cite{fuks2020physics, almajid2022prediction, fraces2021physics}. These limitations highlight the need for more efficient and generalizable approaches to handle the complexity inherent in phase field simulations.

Recent developments in machine learning have also given rise to neural operators, which provide a resolution-invariant framework for solving parametric PDEs \cite{bhattacharya2021model, li2020neural, wang2021learning, eshaghi2024variational}. Unlike traditional data-driven models, neural operators, in both data-driven and physics-based approaches, learn mappings between infinite-dimensional function spaces, allowing them to generalize across different resolutions and domains. Prominent architectures such as Fourier Neural Operators (FNOs) \cite{li2020fourier}, DeepONets \cite{lu2021learning}, Laplace Neural Operators (LNOs) \cite{cao2023lno}, and Convolutional Neural Operators (CNOs) \cite{raonic2024convolutional} have demonstrated superior capabilities in solving complex PDEs, making them particularly well-suited for modeling the intricate dynamics of phase field problems. It is worth noting that neural network frameworks such as those in \cite{khoo2021solving} treat the solution operator as a finite-dimensional regression problem.  In contrast, neural operator methods learn mappings between infinite-dimensional function spaces, enabling them to capture complex functional dependencies across spatial and frequency domains. Furthermore, operator learning fundamentally differs from model order reduction-based methods such as \cite{fresca2022pod}, which first reduce the solution space to a low-dimensional basis and then model solution behavior within this reduced space. 

In most neural operator methods in the literature, such as Phase-Field DeepONet \cite{li2023phase} and FNO-2d \cite{li2020fourier}, the network outputs the solution field for the next time step based on the solution field of the previous steps. Consequently, these methods require iterative evaluation to predict solutions across all time steps. While this approach is sufficient when dealing with small time steps, it becomes problematic in practical scenarios, such as phase field problems, where a large number of time steps are required. In such cases, the iterative process introduces significant cumulative error and poses challenges during training, often exacerbating problems such as the vanishing gradient problem.

Methods such as FNO-3d \cite{li2020fourier} address some of these shortcomings by generating solution fields for all time steps simultaneously. However, this approach has its challenges. Including time steps as an additional dimension (e.g., the third dimension in 3D tensors) requires repeating the 2D solution field \(N_t\) times, where \(N_t\) is the number of time steps. This process results in an inflated 3D tensor, where one dimension is identical across repetitions. This tensor flows through the entire mesh architecture, effectively multiplying all mesh parameters without adding significant new information. As a result, the number of required parameters grows dramatically, making the approach computationally expensive and inefficient. The scalability issue becomes even more critical as \(N_t\) increases.

To resolve these challenges, we use a modified structure similar to Phase-Field DeepONet or FNO-2d and pass the 2D field into the network but at the latest step, for projecting back to the target dimension, we design separate neural networks for each time step instead of employing a single network. These networks collectively allow the neural operator to predict the entire solution for all time steps in a single forward pass, thereby addressing the problem of cumulative error while maintaining computational efficiency.

Another issue with the FNO-3d architecture is the way solution fields for different time steps are generated in the output layer. Each time step corresponds to a separate output channel, and there is no explicit connection between the solution fields at successive time steps. This lack of inter-step connectivity is a significant limitation because, in physical dynamic systems, the solution field at time \(t\) strongly influences the solution field at time \(t+1\). One method for addressing this connectivity issue is to use attention mechanisms, such as those described in \cite{bonneville2024accelerating}. However, attention mechanisms are primarily designed for use in domains like natural language processing, where the relationships between words are complex and not necessarily sequential, allowing later words to influence the meaning of earlier ones. In contrast, physical dynamic systems do not exhibit this bidirectional influence—earlier states govern later states, not the other way around. Moreover, most of the temporal information needed to predict the solution at the next time step is contained in a few preceding time steps. Using attention mechanisms in this context unnecessarily increases model complexity and parameter count, leading to a greater risk of overfitting without significant gains in accuracy.

Our approach to addressing the temporal connectivity issue is inspired by the message-passing paradigm. Specifically, we introduce explicit connections between consecutive time steps. Each time step's output not only provides the solution for that step but also serves as an input to another neural network for the subsequent time step. This design creates a structured dependency across time, enabling the model to capture long-term temporal dynamics naturally. By progressively passing information through the networks for subsequent time steps, our method ensures that the solution evolves consistently over time while efficiently managing dependencies. This approach achieves an effective balance between performance and computational efficiency, addressing the core limitations of both iterative and simultaneous prediction methods. 

In the following sections, we explored the use of the neural operators in more detail, highlighting their applications in various phase-field problems. We first provided an overview of the theoretical foundation of neural operators in Sec. \ref{sec:NeuralOperator}. In Sec. \ref{sec:MHNO}, we introduced MHNO architectures, emphasizing their ability to approximate complex physical systems. Following that, in Sec. \ref{sec:Experiment}, we discussed the MHNO application to phase-field problems and provided numerical results that demonstrate MHNO's capability to handle phase-field problems with high efficiency. We concluded in Sec. \ref{sec:Conclusion} with a summary of key findings and potential directions for future research. The code and data supporting this work are publicly available at \href{https://github.com/eshaghi-ms/MHNO}{https://github.com/eshaghi-ms/MHNO}.

\section{Neural Operator} \label{sec:NeuralOperator}

The neural operator methodology, proposed by Li et al. \cite{li2020neural}, is designed to learn mappings between infinite-dimensional spaces, offering an efficient approach to solving parametric PDEs across varying instances of parameters. Let \(\mathcal{D} \subset \mathbb{R}^d\) be a bounded, open set, and define \(\mathcal{A} = \mathcal{A}(\mathcal{D}; \mathbb{R}^{d_a})\), \(\mathcal{U} = \mathcal{U}(\mathcal{D}; \mathbb{R}^{d_u})\), and \(\mathcal{V} = \mathcal{V}(\mathcal{D}; \mathbb{R}^{d_v})\) as separable Banach spaces of functions taking values in \(\mathbb{R}^{d_a}\) and \(\mathbb{R}^{d_u}\), respectively. Consider \(\mathcal{G}^\dagger: \mathcal{A} \to \mathcal{U}\), a nonlinear solution operator for parametric PDEs. Given \(\mathcal{N}\) observations \(\{a_j, u_j\}_{j=1}^\mathcal{N}\), where \(a_j \sim \mu\) (an i.i.d. sample from the probability measure \(\mu\)) and \(u_j = \mathcal{G}^\dagger(a_j)\), the goal is to approximate \(\mathcal{G}^\dagger\) with a parametric map:  
\begin{equation}
\mathcal{G}: \mathcal{A} \times \Theta \to \mathcal{U} \quad \text{or equivalently,} \quad \mathcal{G}_\theta: \mathcal{A} \to \mathcal{U}, \; \theta \in \Theta,  
\end{equation}
where \(\Theta\) is a finite-dimensional parameter space, and \(\theta^\dagger \in \Theta\) is optimized such that \(\mathcal{G}_\theta \approx \mathcal{G}^\dagger\). 

The neural operator approximates \(\mathcal{G}^\dagger\), as an iterative process \(v_0 \to v_1 \to \dots \to v_\mathcal{T}\), where \(v_j \in \mathcal{V}(\mathcal{D}; \mathbb{R}^{d_v})\) represents a sequence of intermediate representations in a latent space. The forward pass of the model begins by lifting the input \(a \in \mathcal{A}\) to a higher-dimensional representation, \(v_0(x) = \mathcal{P}(a(x))\), where \(\mathcal{P}\) is a local transformation parameterized by a shallow fully connected neural network. Subsequently, the representations undergo \(\mathcal{T}\) iterative updates and each update \(v_{\tau} \to v_{\tau+1}\) is expressed as:  
\begin{equation}
v_{\tau+1}(x) := \sigma \left( \mathcal{W} v_\tau(x) + (\mathcal{K}(a; \phi)v_t)(x) \right), \quad \forall x \in \mathcal{D},
\label{eq:iterative_update}
\end{equation} 
where \(\mathcal{W}: \mathbb{R}^{d_v} \to \mathbb{R}^{d_v}\) is a linear transformation, \(\sigma: \mathbb{R} \to \mathbb{R}\) is a nonlinear activation function applied component-wise, and \(\mathcal{K}: \mathcal{A} \times \Theta_\mathcal{K} \to \mathcal{L}(\mathcal{U}(\mathcal{D}; \mathbb{R}^{d_v}), \mathcal{U}(\mathcal{D}; \mathbb{R}^{d_v}))\) is a bounded linear operator parameterized by \(\phi \in \Theta_\mathcal{K}\). The non-local operator \(\mathcal{K}\) is realized as a kernel integral operator, for which we used the Fourier integral operator \cite{li2020fourier}. Finally, the output \(u(x)=\mathcal{Q}(v_\mathcal{T}(x))\) is the projection of \(v_\mathcal{T}\) by the local transformation \(\mathcal{Q}:\mathbb{R}^{d_v} \mapsto \mathbb{R}^{d_u} \). Therefore, the neural operator is defined as:
\begin{equation}
\mathcal{G}_\theta := \mathcal{Q} \circ (\mathcal{W}_L + \mathcal{K}_L) \circ \cdots \circ \sigma(\mathcal{W}_1 + \mathcal{K}_1) \circ \mathcal{P}
\label{eq:neural_operator}
\end{equation}
A remarkable instance of the neural operator is the Fourier Neural Operator (FNO), which has received considerable attention for its ability to model complex parametric PDEs efficiently. The FNO uses the Fast Fourier Transform (FFT) to represent the non-local kernel operator \(\mathcal{K}\) in the frequency domain, allowing the computation of global dependencies with reduced computational complexity. Specifically, the kernel \(\mathcal{K}(a; \phi)\) is parameterized as a truncated series of Fourier modes, which significantly compresses the information while preserving essential features of the solution operator.

For time-dependent problems, the FNO offers two different implementations, FNO-2d and FNO-3d, which are designed to handle temporal dynamics differently while addressing two-dimensional spatial domains. FNO-2d adopts an RNN structure in time, iteratively updating the solution state step by step as the temporal evolution progresses. Consequently, the network output represents the solution one time step into the future. In contrast, FNO-3d treats space-time as a single three-dimensional domain and performs convolutions directly within this unified representation. This approach allows FNO-3d to capture spatiotemporal correlations comprehensively without explicitly disentangling spatial and temporal computations. The network output in this case provides the solution field over full temporal intervals.

\section{Multi-Head Neural Operator (MHNO)} \label{sec:MHNO}

Multi-Head Neural Operator (MHNO) proposes replacing the global projection operator \(\mathcal{Q}\) in Eq. \ref{eq:neural_operator} with a collection of time-specific neural networks \(\{\mathcal{Q}_n\}_{n=1}^{N_t}\) for each time step \(n\). Additionally, it incorporates a second set of neural networks \(\{\mathcal{H}_n\}_{n=2}^{N_t}\) to explicitly model the connections between time steps, thereby introducing step-wise temporal dependencies into the neural operator architecture. 

\begin{figure}[H]
    \centering
    \includegraphics[width=1.0\textwidth]{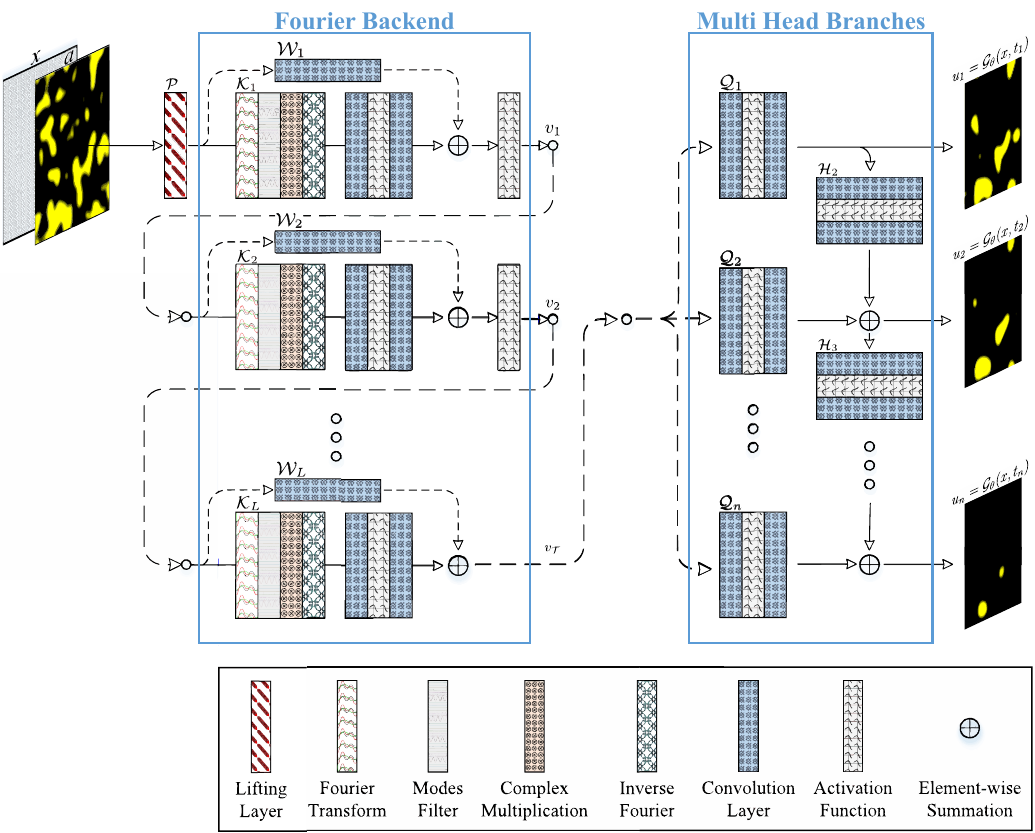}
    \caption{\textbf{Architectures of Multi Head Neural Operator.} Start from input \(a\), 1. Lift to higher dimension channel space by neural network \(\mathcal{P}\), 2. Apply \(L\) layers Fourier operator and activation function, 3. Convert to \(n\) branches by neural networks \(\{\mathcal{Q}_n\}\), 4. Finalize by adding neural networks \(\{\mathcal{H}_n\}\).}
    \label{fig: flowchart}
\end{figure}

In this framework, which is illustrated in Figure \ref{fig: flowchart}, the shared layers form a neural operator, processing the input \(a(x)\) and encoding the global dynamics across the spatial domain. The sequence of intermediate latent variables \(\{v_\tau\}_{\tau=0}^\mathcal{T}\), computed through the iterative updates of the neural operator, represents the higher-dimensional transformation of the input. The terminal representation \(v_\mathcal{T}(x)\), instead of being projected directly into the output space by a single global operator \(\mathcal{Q}\), is passed through the step-specific projection networks \(\mathcal{Q}_n\). Each \(\mathcal{Q}_n\) specializes in producing the solution field \(u_n(x)\) for its corresponding time step \(n\), ensuring that the solution fields reflect the underlying spatial dynamics as well as the temporal dependencies embedded within the input and intermediate representations.

At its core, the revised methodology modifies the standard projection step in neural operators to enable direct interaction between adjacent time steps. Therefore, the final output at time \(t_n\) corresponding to time step \(n\) is expressed as:  
\begin{equation}
\mathcal{G}_\theta(x,t_n)(a(x)) := \mathcal{Q}_n \circ (\mathcal{W}_L + \mathcal{K}_L) \circ \cdots \circ \sigma(\mathcal{W}_1 + \mathcal{K}_1) \circ \mathcal{P} (a(x)) + \mathcal{H}_n \circ \mathcal{G}_\theta(x,t_{n-1})(a(x))
\label{eq:MHNO}
\end{equation}
 
where \(\mathcal{G}_\theta(x,t_{0})\) is the identity operator \(I(x)\), \(\mathcal{H}_1\) is the zero operator \(\mathcal{O}(x)\), \(\mathcal{Q}_n\) projects the intermediate latent representation \(v_\mathcal{T}\) at step \(n\), and \(\mathcal{H}_n \circ \mathcal{G}_\theta(x,t_{n-1})\) integrates information from the preceding time step \(n-1\) into the current prediction. This structure introduces a temporal connection mechanism inspired by the principles of message passing, where outputs from previous time steps directly influence subsequent predictions.

The use of independent neural networks \(\{\mathcal{Q}_n\}\) and \(\{\mathcal{H}_n\}\) allows MHNO to adapt dynamically to the requirements of each time step. Unlike global operators like \(\mathcal{Q}\) in standard neural operators, the step-specific projection operators \(\mathcal{Q}_n\) are tailored to the characteristics of individual time steps, enhancing their ability to model temporal variation effectively. Similarly, the inclusion of \(\mathcal{H}_n\) ensures that each step receives information about the evolving temporal context, capturing long-term dependencies while maintaining computational efficiency.

Let us define a restricted subclass of MHNOs where all temporal connections are removed by setting, $\mathcal{H}_n = \mathcal{O}$ (the zero operator), then MHNO reduces to a collection of independent neural operators with time-specific output heads $\{\mathcal{Q}_n\}$:
\begin{equation}
    \mathcal{G}_\theta(x,t_n)(a(x)) := \mathcal{Q}_n \circ (\mathcal{W}_L + \mathcal{K}_L) \circ \cdots \circ \sigma(\mathcal{W}_1 + \mathcal{K}_1) \circ \mathcal{P} (a(x)), 
\end{equation}
which is exactly the form of a standard neural operator with step-specific output projections. In particular, if all $\mathcal{Q}_n = \mathcal{Q}$, it exactly matches the standard neural operator formulation. Therefore, standard neural operators are a strict subclass of MHNOs, and we know that the approximation theorem is true for standard neural operators (\cite{kovachki2023neural}, Theorem 11). Since this restricted version of MHNO is a subset of the full MHNO class and already satisfies approximation theorem, it follows that the full MHNO class also satisfies the same approximation guarantee.

\begin{theorem}
Let Assumptions 9 and 10 in \cite{kovachki2023neural} hold, and suppose $\mathcal{G}^\dagger : \mathcal{A} \to \mathcal{U}$ is continuous. Then for any compact set $\mathcal{K} \subset \mathcal{A}$ and any $0 < \varepsilon \leq 1$, there exists an $\mathcal{G}_\theta$, based on the MHNO architecture, such that
\begin{equation}
    \sup_{a \in \mathcal{K}} \|\mathcal{G}^\dagger(a) - \mathcal{G}_\theta(a)\|_\mathcal{U} \leq \varepsilon.
\end{equation}

Furthermore if $\mathcal{U}$ is a Hilbert space, and the target operator satisfies $\|\mathcal{G}^\dagger(a)\|_\mathcal{U} \leq M$ for all $a \in \mathcal{A}$ and some $M>0$, and if the output operators $\mathcal{Q}_n$ have bounded norm, $\|\mathcal{Q}_n\| \leq 1$, the temporal coupling operators $\mathcal{H}_n$ satisfy $\|\mathcal{H}_n\| \leq \gamma < 1$, then $\mathcal{G}_\theta(a; t_n)$ satisfy the following bound:
\begin{equation}
    \|\mathcal{G}_\theta(a; t_n)\|_\mathcal{U} \leq 4M, \quad \forall a \in \mathcal{A}.
\end{equation}
\end{theorem}

This revised architecture provides several advantages over existing approaches like FNO-2d, Phase-Field DeepONet, and FNO-3d. By predicting all time steps in a single forward pass and maintaining explicit connections between them, MHNO mitigates the challenges of error accumulation and parameter inflation. The explicit dependency between time steps enables better modeling of dynamic systems, where earlier states significantly influence subsequent outcomes. Furthermore, the modularity of MHNO’s step-specific components makes it scalable to problems involving long time horizons, avoiding the parameter explosion associated with methods like FNO-3d.

\begin{figure}[H]
    \centering
    \includegraphics[width=0.7\textwidth]{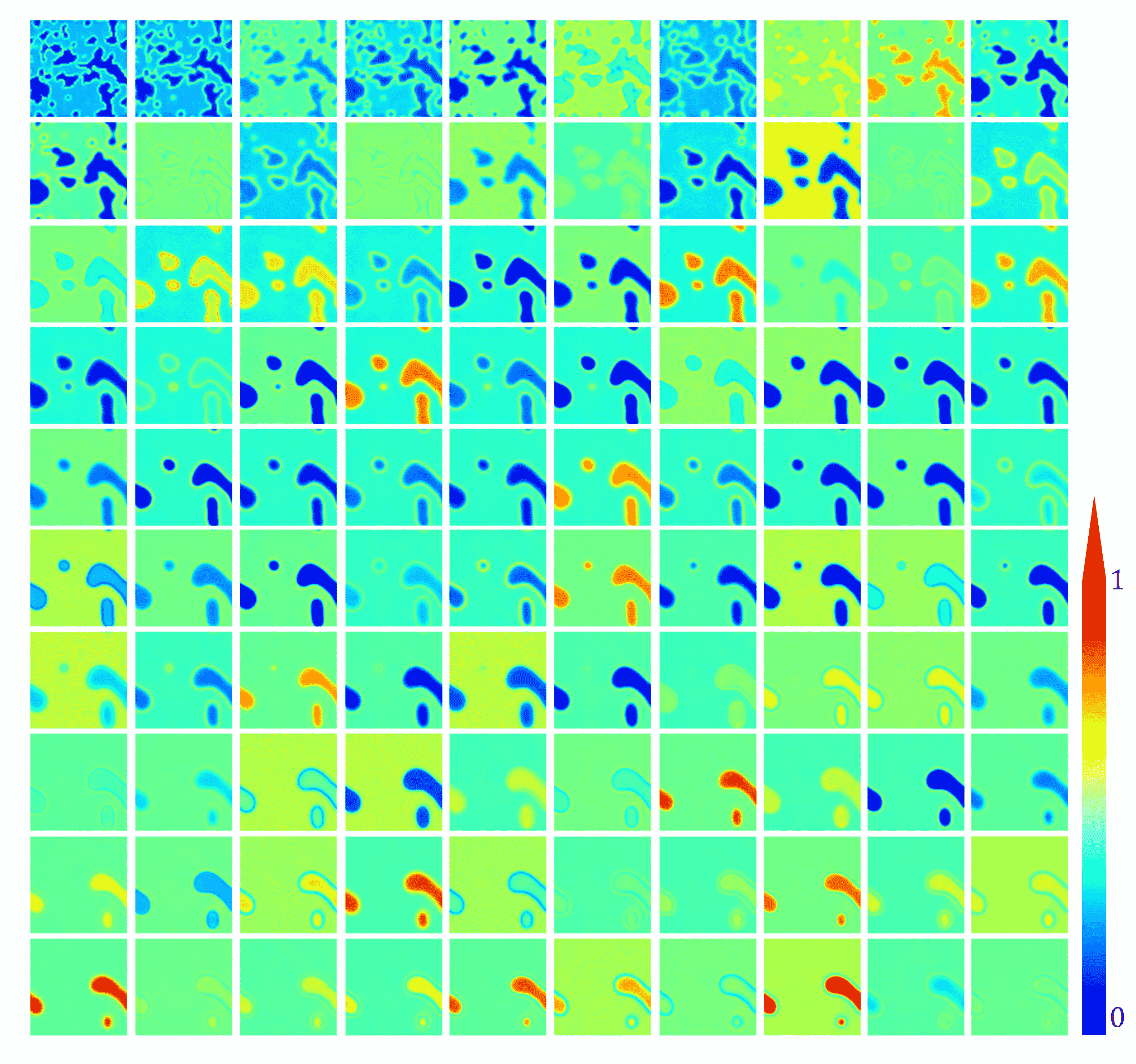}
    \caption{\(\mathcal{H}_n\) contributions over time, showing the evolving ratio \(R_H^{(n)}\) that highlights the dynamic interplay between the two terms, \(\mathcal{H}_n\) and \(\mathcal{Q}_n\). The periodic pattern indicates their changing importance throughout the time steps. The plot starts from the first time step at the top left of the figure and progresses to the 100th time step at the bottom right.}
    \label{fig: Q_H}
\end{figure}

To better understand the contributions of \(\mathcal{H}_n\) and \(\mathcal{Q}_n\) to the final solution, we compared their relative influence using the metric
\begin{equation}
R_H^{(n)} = \frac{\exp(\lvert H_n \rvert)}{\exp(\lvert Q_n \rvert) + \exp(\lvert H_n \rvert)},
\end{equation}
which serves as a normalized index of the importance of the contributions of \(\mathcal{H}_n\) at each time step. We computed \(R_H^{(n)}\) values for 100 time steps in solving the Allen-Cahn equation, and Figure \ref{fig: Q_H} illustrates their variation over time, highlighting the changing importance of the temporal dependencies introduced by \(\mathcal{H}_n\). The figure shows that both \(\mathcal{H}_n\) and \(\mathcal{Q}_n\) make substantial contributions, with their ratio exhibiting a periodic pattern. This emphasizes the dynamic interaction between \(\mathcal{H}_n\) and \(\mathcal{Q}_n\) over time and confirms their essential role in the final solution.

\section{Experiments} \label{sec:Experiment}

To evaluate the effectiveness and generality of the proposed Multi-Head Neural Operator (MHNO), we conducted a comprehensive set of experiments on a diverse range of phase field models, each governed by distinct physical principles and associated PDEs. The selected benchmarks encompass key phenomena encountered in materials science and nonlinear dynamics, including antiphase boundary motion, spinodal decomposition, pattern formation, atomic scale modeling, and molecular beam epitaxy growth model. 

In our experiments, we compare the performance of MHNO against prominent neural operator methods, Fourier Neural Operator with an RNN structure in time (FNO-2d), and those that directly convolves in space-time (FNO-3d). The comparison is based on several metrics, including accuracy, computational efficiency, parameter number (memory usage), and the ability to capture long-term dynamics of the solution fields. These metrics are particularly crucial in practical applications where maintaining numerical precision while handling large numbers of time steps is essential.  

For each problem, we generated a high-fidelity dataset by numerically solving the PDE on a periodic domain using a spectral method. Time integration was performed with a sufficiently small solver time step (denoted here as $\Delta t_s$) to ensure that the “ground truth” trajectories are well resolved. From each trajectory, we then extracted $N_t+1$ discrete snapshots at uniform intervals (denoted as the \emph{network time step}, $\Delta t$), which differs from $\Delta t_s$. Each trajectory thus consists of $N_t+1$ fields $\{\mathbf{\phi}^0,\,\mathbf{\phi}^1,\,\dots,\,\mathbf{\phi}^{N_t}\},$ where $\mathbf{\phi}^0$ is the initial condition and $\{\mathbf{\phi}^1,\dots,\mathbf{\phi}^{N_t}\}$ are the solution fields at the subsequent $N_t$ time steps. All networks are trained on these snapshot sequences, and the goal is to predict the $N_t$ future fields from the initial condition.

We have considered three different approaches to predict the entire sequence:

\begin{enumerate}
    \item \textbf{Approach 1 (Windowed Subsequence Training).} 
    We partition each trajectory of $N_t$ intervals into windows of length $n_w$, and we extract \emph{all} possible subsequences of $n_w$ intervals. Each subsequence spans $n_w$ intervals and thus $n_w+1$ snapshots. Concretely, for each valid starting index $k=0,1,\dots,N_t - n_w$, we construct a training sample $\bigl(\mathbf{\phi}^{\,k},\;\mathbf{\phi}^{\,k+1},\;\dots,\;\mathbf{\phi}^{\,k + n_w}\bigr)$. Thus, the total number of such samples per trajectory is $N_t - n_w + 1$. In each sample, the network learns to map the first snapshot $\mathbf{\phi}^{\,k}$ to the next $n_w$ snapshots $(\mathbf{\phi}^{\,k+1},\dots,\mathbf{\phi}^{\,k + n_w})$. During inference over the full trajectory, we proceed sequentially in windows of size $n_w$, chaining predictions as follows: starting from $\mathbf{\phi}^0$, we predict $\mathbf{\phi}^1,\dots,\mathbf{\phi}^{n_w}$; then using the predicted $\mathbf{\phi}^{\,n_w}$ we predict $\mathbf{\phi}^{\,n_w+1},\dots,\mathbf{\phi}^{2n_w}$; and we continue in this fashion until obtaining $\{\mathbf{\phi}^1,\dots,\mathbf{\phi}^{N_t}\}$. Although training samples overlap in time, inference remains windowed and non-overlapping. In our experiments, we have selected \(n_w=10\).
    
    \item \textbf{Approach 2 (Two-Window Training).} 
    Here, we set $n_w = N_t/2$ (assumed integer) and again extract \emph{all} possible subsequences of length $n_w$ intervals, i.e., for each $k=0,1,\dots,N_t - n_w$,  we construct $\bigl(\mathbf{\phi}^{\,k}, \,\mathbf{\phi}^{\,k+1},\,\dots,\,\mathbf{\phi}^{\,k + n_w}\bigr)$. In inference, exactly two non-overlapping windows of size $n_w$ intervals each have been used. First predict $\mathbf{\phi}^1,\dots,\mathbf{\phi}^{n_w}$ from $\mathbf{\phi}^0$, and then we use the predicted $\mathbf{\phi}^{\,n_w}$ to predict $\mathbf{\phi}^{\,n_w+1},\dots,\mathbf{\phi}^{\,N_t}$. Although training samples overlap, inference only requires two windows.
    
    \item \textbf{Approach 3 (Full-Trajectory Training).} 
    In the last approach, we used the entire set of $N_t+1$ snapshots as a single training sample, in which input is $\mathbf{\phi}^0$ and output is $(\mathbf{\phi}^1,\mathbf{\phi}^2,\dots,\mathbf{\phi}^{\,N_t})$. Thus, no temporal partitioning is performed, so the network learns the direct mapping $\mathbf{\phi}^0 \;\longmapsto\; \bigl(\mathbf{\phi}^1,\,\mathbf{\phi}^2, \,\dots,\,\mathbf{\phi}^{\,N_t}\bigr)$. Also, during inference, one forward pass predicts all $N_t$ future snapshots. 
\end{enumerate}

In all approaches, we maintain the same number of trajectories and the same spatial resolution. The only difference lies in how the $N_t+1$ snapshots are grouped into training samples.

To create the database used in these experiments, we solve these PDEs using the Fourier spectral method \cite{yoon2020fourier}, which provides high-accuracy solutions suitable for training and evaluation. For each PDE, we generated \(N_{\text{train}} + N_{\text{test}}\) random initial conditions, using a Gaussian random field, on a uniform grid of size $N_x \times N_y$ with periodic boundary conditions. These initial conditions differ substantially from those used in studies such as \cite{zhang2024energy}, which rely solely on sine functions as input. Such restricted inputs tend to yield artificially high accuracy due to similarity within the dataset.  Time integration used a spectral discretization in space and a semi-implicit time stepping in time, with $\Delta t_s$ chosen to be sufficiently small to ensure numerical stability and accuracy. From each fully resolved trajectory $\{\mathbf{\phi}(t)\}_{t=0}^{T}\}$, we sampled $N_t+1$ uniformly spaced “network time” snapshots at intervals $\Delta t = T / N_t$. The computational times for the neural operator have been obtained using the PyTorch library on an NVIDIA A100-PCIE-40GB GPU. Detailed information regarding implementing the Fourier spectral method can be found in Appendix \ref{appendix:A}. 

\subsection{Antiphase Boundary Motion - Allen-Cahn Equation}

We address the problem of antiphase domain coarsening, where spatially ordered regions in alloys grow and merge over time due to the motion of antiphase boundaries. These boundaries separate domains with different atomic arrangements and tend to migrate to reduce the total interfacial energy, leading to a coarsening process characterized by the elimination of small domains and the growth of larger ones. The theoretical foundation for this phenomenon was established by Allen and Cahn, who modeled the motion of these boundaries using a curvature-driven mechanism. They showed that the interfacial velocity is proportional to the local mean curvature and that the driving force for coarsening arises from diffusional dissipation, not from the magnitude of the specific surface free energy. This framework results in a phase-field formulation commonly described by a reaction-diffusion equation involving a double-well potential, known in the literature as the Allen-Cahn equation.

The Allen-Cahn equation is a widely studied PDE used to model phase separation and interface motion in a system with two distinct phases. The Allen-Cahn equation, with a wide range of applications including two-phase fluids \cite{chen2023mean}, complex dynamics of dendritic growth \cite{xiao2022second}, image inpainting \cite{dobrosotskaya2008wavelet}, etc, is computationally challenging to solve because it requires fine spatial and temporal discretizations to capture the evolution of the thin interface regions accurately. This makes it a suitable benchmark problem for evaluating the performance of neural operator-based solvers in resolving phase field dynamics.

\begin{figure}[H]
    \captionsetup{skip=-5pt}
    \centering
    \includegraphics[width=0.95\textwidth]{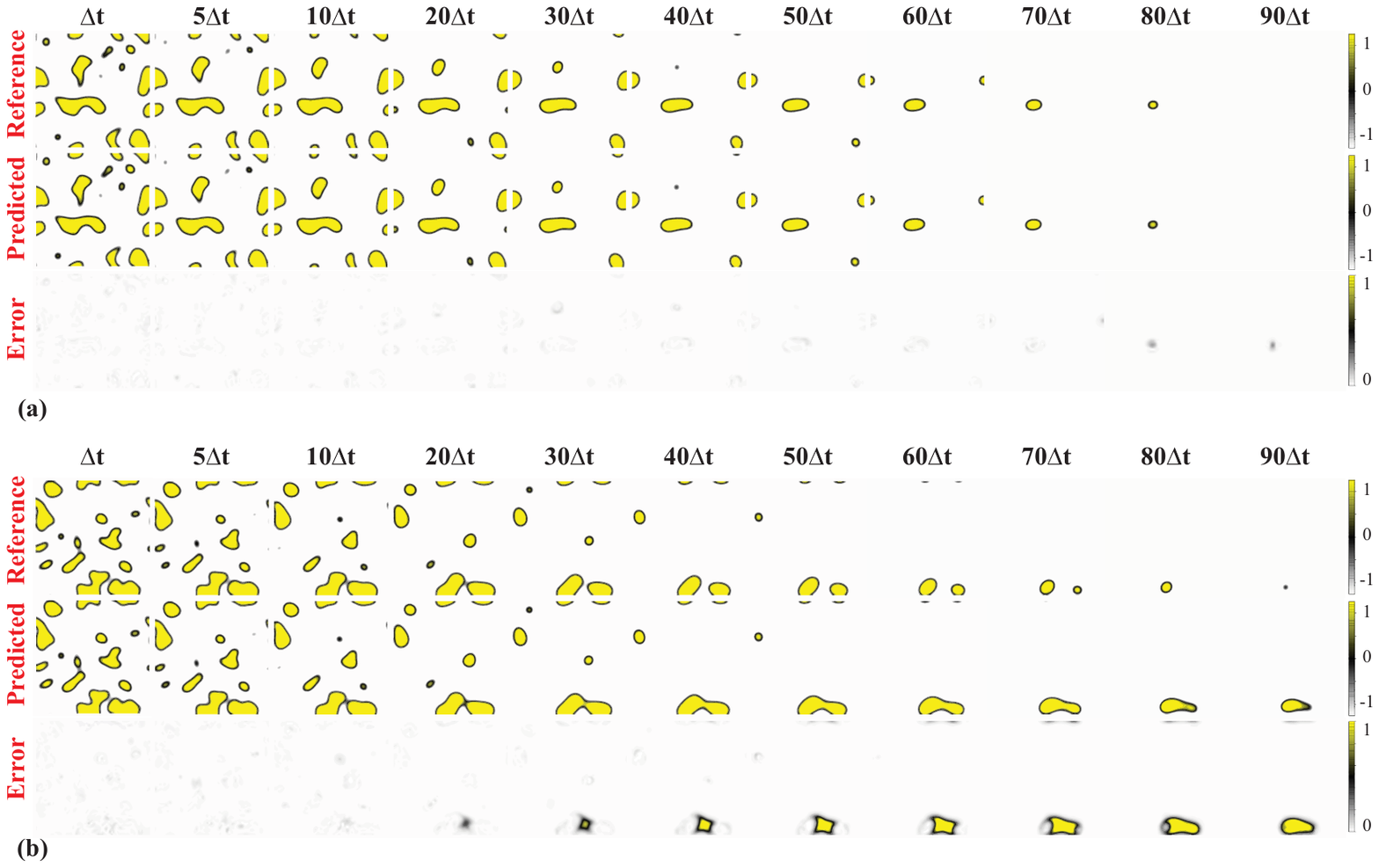}
    \caption{Phase-field approximation of the antiphase boundary motion from different initial conditions. The shapes are represented by the phase-field variable \( \phi \), where \( 0 \leq \phi \leq 1 \).}
    \label{fig: AC2d-different-BC}
\end{figure}

The Allen-Cahn equation is derived from minimizing a free energy functional, where the system evolves over time to reduce its total energy. We consider the 2D Allen-Cahn equation for a phase field variable \( \phi(x, t) \) \cite{allen1979microscopic}:  
\begin{equation}
\frac{\partial \phi(x, t)}{\partial t} = \Delta \phi(x, t) - \frac{1}{\epsilon^2} f'(\phi(x, t)), \quad x \in [-\pi, \pi]^2, \; t \in (0, T],
\end{equation}
\begin{equation}
\phi(x, 0) = \phi_0(x), \quad x \in [-\pi, \pi]^2,
\end{equation}

where \( \epsilon > 0 \) is a parameter controlling the interface width, \( f(\phi) = \frac{1}{4}(\phi^2 - 1)^2 \) is a double-well potential, and \( f'(\phi) \) is its first derivative. The Laplace operator \( \Delta\phi(x, t) \) governs diffusion, while \( f'(\phi) \) enforces the separation into the two equilibrium phases \( \phi = \pm 1 \). The initial condition \( \phi_0(x) \) defines the phase field distribution at \( t = 0 \). We are interested in approximating the operator that maps the initial state \( \phi(x, 0) \) to the phase field's temporal evolution over the interval \( (0, T] \), defined as \(\mathcal{G}^\dagger: \phi_0(x) \mapsto \phi(x, t) \quad \text{for } t \in (0, T].\) This involves predicting the entire spatiotemporal evolution of the system from its initial configuration. For the experiments, we set \( \epsilon = 0.05 \), $\Delta t_s = 0.01$ and the number of time steps \( N_t = 90 \) and \(\Delta t = 1\) and simulate the dynamics on a fixed resolution of \( 64 \times 64 \) for both training and testing. Examples of the evolution of the phase ﬁeld variable, \( \phi(x, t) \), as a function of time for different initial conditions are presented in Figure \ref{fig: AC2d-different-BC}.

\begin{table}[H]
    \centering
    \caption{Performance comparison of neural operator models for solving the anthiphase boundary motion across different approaches}
    \label{tab:AC2D}
    \resizebox{\textwidth}{!}{%
    \begin{tabular}{l|lll|lll|lll|lll|lll}
    \toprule
    \multirow{2}{*}{\begin{tabular}[c]{@{}l@{}}\textbf{Temp.}\\ \textbf{Appr.}\end{tabular}}
    & \multicolumn{3}{c|}{\textbf{Parameters}}
    & \multicolumn{3}{c|}{\textbf{Training Time (s/epoch)}}
    & \multicolumn{3}{c|}{\textbf{Inference Time (s)}}
    & \multicolumn{3}{c|}{\textbf{Train $L^2$-error (\%)}}
    & \multicolumn{3}{c}{\textbf{Test $L^2$-error (\%)}}
    \\ \cline{2-16}
    & 
    \multicolumn{1}{c}{MHNO} & \multicolumn{1}{c}{FNO-3d} & \multicolumn{1}{c|}{FNO-2d} & 
    \multicolumn{1}{c}{MHNO} & \multicolumn{1}{c}{FNO-3d} & \multicolumn{1}{c|}{FNO-2d} & 
    \multicolumn{1}{c}{MHNO} & \multicolumn{1}{c}{FNO-3d} & \multicolumn{1}{c|}{FNO-2d} & 
    \multicolumn{1}{c}{MHNO} & \multicolumn{1}{c}{FNO-3d} & \multicolumn{1}{c|}{FNO-2d} & 
    \multicolumn{1}{c}{MHNO} & \multicolumn{1}{c}{FNO-3d} & \multicolumn{1}{c}{FNO-2d} \\ \hline
    I   & 2,383,211                     & 4,197,937                     & 4,208,193 
        & \multicolumn{1}{c}{3.50}      & \multicolumn{1}{c}{17.27}     & \multicolumn{1}{c|}{16.00} 
        & \multicolumn{1}{c}{0.0475}    & \multicolumn{1}{c}{0.0408}    & \multicolumn{1}{c|}{0.2877} 
        & $0.24 \pm 0.07$               & $0.58 \pm 0.20$               & $0.53 \pm 0.33$
        & $2.91 \pm 3.11$               & $4.27 \pm 3.73$               & $7.13 \pm 4.52$ \\ 
        
    II  & 2,422,201                 & 4,197,937                         & 4,208,193
        & \multicolumn{1}{c}{12.21}   & \multicolumn{1}{c}{49.34}       & \multicolumn{1}{c|}{58.36} 
        & \multicolumn{1}{c}{0.0336}   & \multicolumn{1}{c}{0.0185}     & \multicolumn{1}{c|}{0.2066} 
        & $0.49 \pm 0.10$               & $0.71 \pm 0.14$               & $1.41 \pm 0.44$
        & $6.84 \pm 4.70$               & $7.62 \pm 5.11$               & $7.75 \pm 3.95$ \\ 
        
    III & 2,472,331                 & 4,197,937                         & 4,208,193 
        &\multicolumn{1}{c}{6.18}   & \multicolumn{1}{c}{30.57}         & \multicolumn{1}{c|}{55.74} 
        &\multicolumn{1}{c}{0.0321} & \multicolumn{1}{c}{0.0157}        & \multicolumn{1}{c|}{0.2009} 
        & $0.73 \pm 0.19$               & $1.29 \pm 0.28$               & $8.91 \pm 1.79$
        & $\textbf{1.89} \pm \textbf{2.09}$ & $3.03 \pm 2.06$           & $11.76 \pm 3.93$\\
    \bottomrule
    \end{tabular}%
    }
\end{table}

Table \ref{tab:AC2D} compares the performance of three neural operator models, MHNO, FNO-3d, and FNO-2d, in solving the 2D Allen-Cahn equation. The models are evaluated based on the number of parameters, average training time per epoch, inference time, and \(L^2\)-error, across different temporal approaches. The number of parameters reflects the model's memory requirement, the training time per epoch indicates computational demand for training, inference time shows the time for predicting the result for the entire period, and the \(L^2\)-error measures prediction accuracy relative to the ground truth.

MHNO demonstrates superior computational efficiency, with the shortest training time per epoch across all configurations. Its time scales reasonably with increasing number of time steps, making it well-suited for large-scale simulations. Additionally, MHNO achieves the best accuracy for all three approaches. FNO-3d and FNO-2d, on the other hand, have a larger number of parameters. In addition, FNO-2d struggles with accuracy, especially when using the third approach. Its training time per epoch, the inference time, and, therefore, its computational cost are higher than the other methods. Additionally, its \(L^2\)-error increases significantly, becoming more than 50 percent at the third approach. Overall, MHNO offers the best balance of efficiency and accuracy for all the approaches, and the result shows that the third approach is a better way to tackle the antiphase boundary motion problem. FNO-3d and FNO-2d, by contrast, show limitations in handling the demands of phase field dynamics at larger time scales.

\begin{figure}[H]
    \centering
    \includegraphics[width=1.0\textwidth]{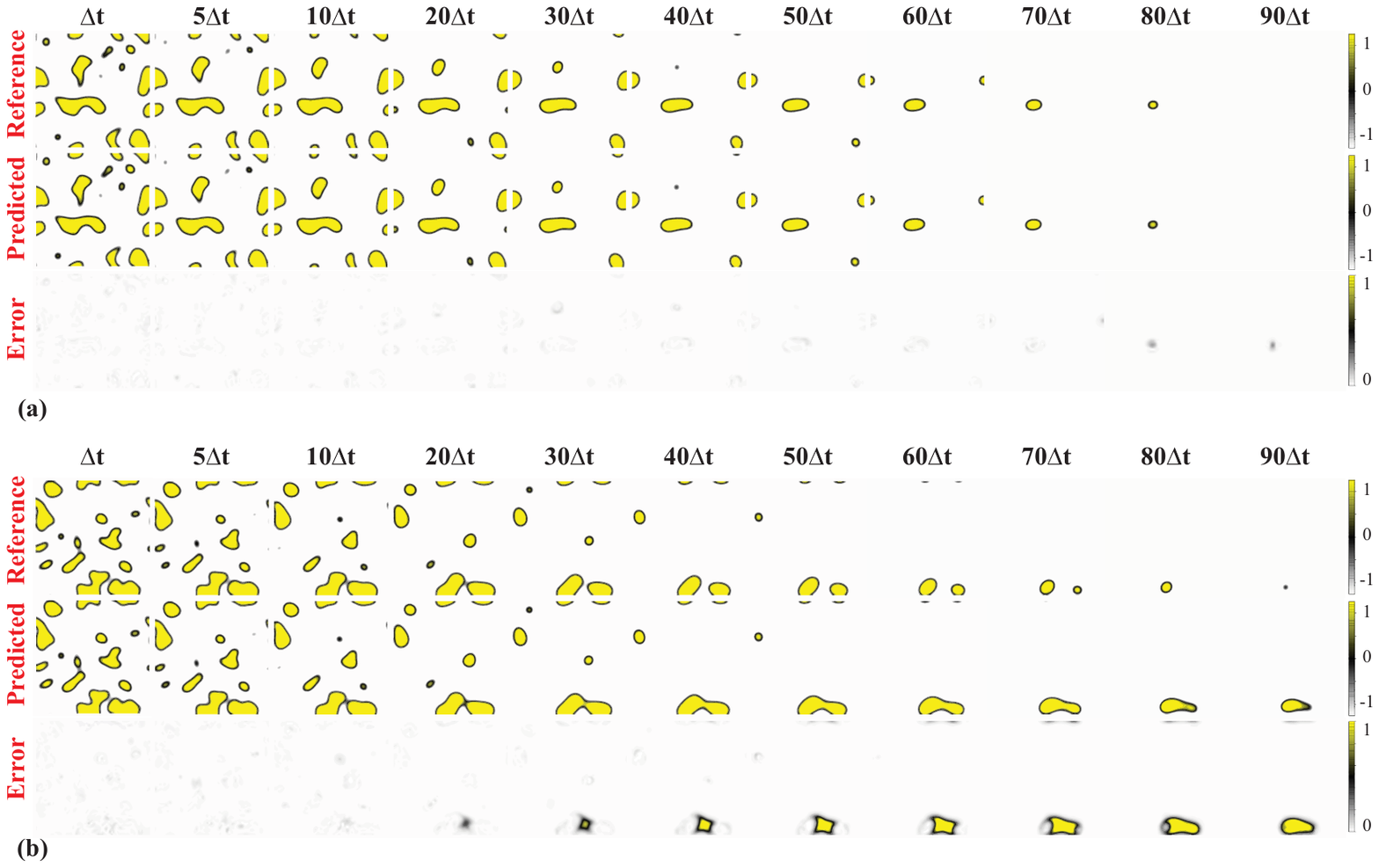}
    \caption{\textbf{Phase Field Predictions and Errors for 2D Allen-Cahn Equation.} Visualization of the reference, predicted, and error fields for the 2D Allen-Cahn equation over time steps (\(\Delta t, 5\Delta t, 10\Delta t, \ldots, 90\Delta t\)). (a) Results for a sample with an error close to the \textbf{mean} \(L^2\) error of the test dataset, showing accurate predictions and minimal discrepancies between the reference and predicted fields. (b) Results for the sample with the \textbf{highest} \(L^2\) error, highlighting regions where the model struggles to accurately capture the phase field dynamics, especially at later time steps. Error plots (bottom rows) highlight the regions of deviation.}
    \label{fig: AC2D}
\end{figure}

Figure \ref{fig: AC2D} shows the reference, predicted, and error fields for the 2D Allen-Cahn equation at different time steps (\(\Delta t, 5\Delta t, 10\Delta t, \ldots, 90\Delta t\)). In panel (a), the results show a sample from the test dataset with an error close to the mean \(L^2\) error, demonstrating accurate predictions by the model. The reference and predicted fields match well across all time steps, with the error fields (bottom row) showing minimal discrepancies between them, indicating that the model effectively captures the phase field evolution. It is worth noting that most samples from the dataset exhibit similar errors to those in the first plot, with the model providing accurate predictions across all time steps. In panel (b), the figure shows the results for a sample with the highest \(L^2\) error, highlighting the challenges the model faces in accurately predicting the phase field dynamics. As the time steps increase, especially for larger \(t\), significant deviations between the reference and predicted fields become more apparent and are visible in the error plots. This second sample can be considered an outlier, as it has a much larger error than most of the test samples, highlighting the variability in prediction accuracy across different samples.

\begin{figure}[H]
    \centering
    \includegraphics[width=1.0\textwidth]{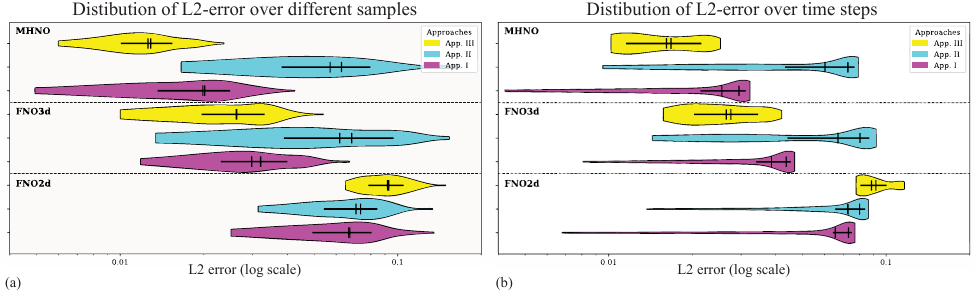}
    \caption{\textbf{$L^2$ Error Distribution for 2D Allen-Cahn Equation.} (a) $L^2$ error distribution across different samples, showing variability in model accuracy. (b) $L^2$ error distribution over time steps, highlighting temporal evolution of prediction errors.}
    \label{fig: AC2D-Comparision}
\end{figure}

To further analyze the performance of the neural operator models, Figure \ref{fig: AC2D-Comparision} presents the distribution of the $L^2$ error across two aspect. Figure \ref{fig: AC2D-Comparision}(a) illustrates the $L^2$ error distribution over different samples in the test dataset, providing insight into the variability of the model's prediction accuracy across different initial conditions, visualizing the superiority of MHNO. Figure \ref{fig: AC2D-Comparision}(b) shows the $L^2$ error distribution over different time steps, revealing how the model's accuracy evolves over the temporal domain. A tightly clustered distribution in this sub-figure would indicate consistent model performance across different time steps, while a broader spread suggests varying accuracy, showing the third approach is more reliable than the others. Together, these distributions complement the results in Table \ref{tab:AC2D} and Figure \ref{fig: AC2D}, offering a comprehensive view of the model's performance across both sample variability and temporal evolution. 

For Allen-Cahn 3D, we considered only the MHNO method and did not consider FNO due to its high computational demand. The 3D Allen-Cahn equation extends the 2D version into three spatial dimensions while maintaining the same governing dynamics of phase separation and interface evolution. Due to the increased dimensionality of the problem, solving the 3D Allen-Cahn equation is more computationally intensive. However, the MHNO model provides an efficient solution framework. The model contains about 16.9 million parameters, and each training iteration takes about 33 seconds. We train the model with a resolution of \( 32^3 \) grid points and simulate the phase evolution over time with parameters (\( \Delta t = 0.01 \), \( T = 1 \)). The results for the 3D simulations similarly demonstrate the effectiveness of the MHNO approach in predicting phase field evolution, with accuracy remaining strong despite the higher dimensionality.

\begin{figure}[H]
    \centering
    \includegraphics[width=1.0\textwidth]{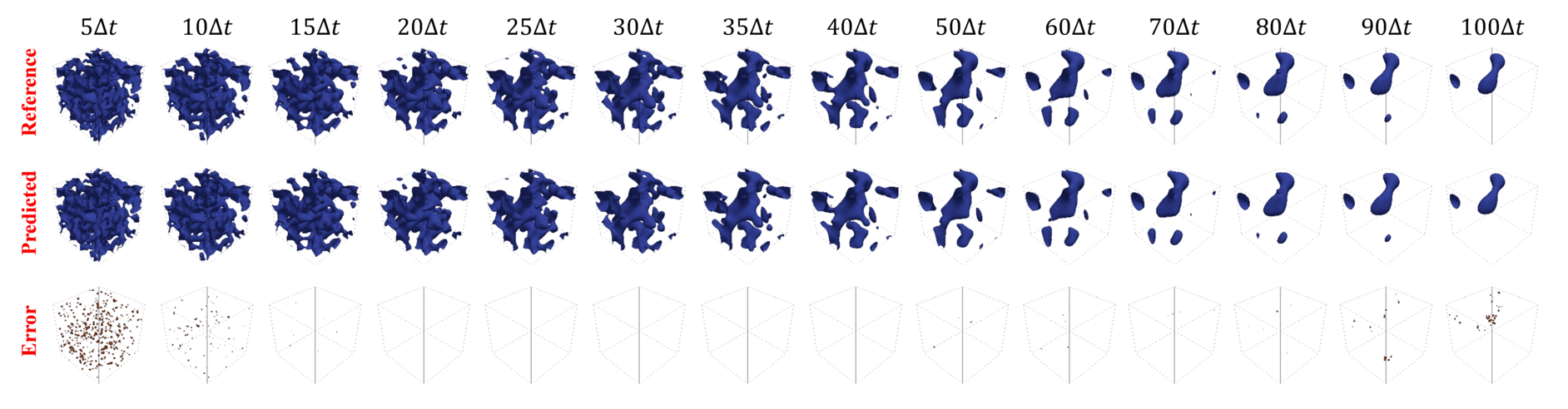}
    \caption{\textbf{Phase Field Predictions and Errors for the 3D Allen-Cahn Equation.} Visualization of the reference, predicted, and error fields for the 3D Allen-Cahn equation over time steps (\(5\Delta t, 10\Delta t, \ldots, 100\Delta t\)) for a sample with an error close to the \textbf{mean} \(L^2\) error of the test dataset, demonstrating the ability of the model to accurately predict the evolution of the phase field. The reference and predicted fields are closely aligned over all time steps, while the error fields (bottom row) show minimal deviations, confirming the robustness of the predictions.}
    \label{fig: AC3D}
\end{figure}

Figure \ref{fig: AC3D} illustrates the performance of the model in predicting the 3D Allen-Cahn equation over a range of time steps (\(5\Delta t, 10\Delta t, \ldots, 100\Delta t\)). A representative test sample with an error close to the mean \(L^2\) error is shown, demonstrating accurate predictions of the phase field dynamics. The reference and predicted fields show high agreement across all time steps, while the error fields (bottom row) show minimal discrepancies, demonstrating the ability of the model to capture the temporal evolution of the 3D phase field. Almost all samples from the test dataset show similarly accurate predictions, as the error for the test dataset is \(1.91 \pm 0.09 \%\) with the minimum value \(1.76 \%\) and maximum value \(2.37 \%\).  

\subsection{Spinodal Decomposition - Cahn–Hilliard Equation}

The second example is the spinodal decomposition problem, characterized by the spontaneous separation of a binary mixture into two distinct phases under mass conservation. The problem is modeled using the Cahn–Hilliard equation, which is another fundamental PDE extensively used to model phase separation and coarsening phenomena in binary systems \cite{cahn1958free}. This problem is governed by the evolution of a conserved order parameter that captures the concentration difference between components and evolves to minimize a free energy functional subject to conservation laws. The dynamics are captured by a fourth-order nonlinear PDE, making traditional numerical simulations computationally intensive, especially at high resolutions needed to resolve thin interfaces.

Unlike the Allen-Cahn equation, which is a non-conserving equation, the Cahn–Hilliard equation describes the evolution of a conserved order parameter, making it particularly suitable for systems where mass conservation is required, such as binary alloys \cite{abazari2022numerical}, polymer blends \cite{inguva2021continuum}, and spinodal decomposition \cite{konig2021two}. The Cahn–Hilliard equation is also derived from a free energy functional, but it incorporates a conservation law that governs the evolution of the phase field variable \( \phi(x, t) \), where \( \phi \) represents the concentration difference between the two phases. The equation has a fourth-order spatial derivative, which introduces additional computational challenges, particularly when resolving sharp interfaces. This makes the Cahn–Hilliard equation a robust benchmark for testing the capabilities of neural operator-based solvers in capturing the complex dynamics of phase separation.

We consider the 2D Cahn–Hilliard equation for the phase field variable \( \phi(x, t) \) \cite{cahn1958free}:  
\begin{equation}
\frac{\partial \phi(x, t)}{\partial t} = \Delta \mu(x, t), \quad \mu(x, t) = -\epsilon^2 \Delta \phi(x, t) + f'(\phi(x, t)), \quad x \in [-0.5, 0.5]^2, \; t \in (0, T],
\end{equation}
\begin{equation}
\phi(x, 0) = \phi_0(x), \quad x \in [-0.5, 0.5]^2,
\end{equation}

where \( \mu(x, t) \) is the chemical potential, \( \epsilon > 0 \) controls the interface width, and \( f(\phi) = \frac{1}{4}(\phi^2 - 1)^2 \) is the double-well potential with \( f'(\phi) \) as its derivative. The term \( -\epsilon^2 \Delta \phi(x, t) \) represents the interfacial energy contribution, while \( f'(\phi) \) enforces phase separation into the two equilibrium states \( \phi = \pm 1 \). 

The conservation of \( \phi \) ensures that the total concentration remains constant throughout the evolution, distinguishing it from the Allen-Cahn equation. The initial condition \( \phi_0(x) \) specifies the phase field distribution at \( t = 0 \). We aim to learn the operator \( G^\dagger: \phi_0(x) \mapsto \phi(x, t) \; \text{for } t \in (0, T] \), mapping the initial configuration to the spatiotemporal evolution of the phase field. For experiments, we use \( \epsilon = 0.0125 \), \( \Delta t_s = 0.005 \) and simulate the system over \( N_t = 90 \) time steps with \( \Delta t = 0.5 \). The dynamics are computed on a fixed resolution of \( 64 \times 64 \) for both training and testing, providing a challenging dataset for evaluating neural operators in resolving conserved phase field dynamics.

\begin{table}[H]
    \centering
    \caption{Performance comparison of neural operator models for solving the spinodal decomposition problem across different approaches.}
    \label{tab:CH2D}
    \resizebox{\textwidth}{!}{%
    \begin{tabular}{l|lll|lll|lll|lll|lll}
    \toprule
    \multirow{2}{*}{\begin{tabular}[c]{@{}l@{}}\textbf{Temp.}\\ \textbf{Appr.}\end{tabular}}
    & \multicolumn{3}{c|}{\textbf{Parameters}}
    & \multicolumn{3}{c|}{\textbf{Training Time (s/epoch)}}
    & \multicolumn{3}{c|}{\textbf{Inference Time (s)}}
    & \multicolumn{3}{c|}{\textbf{Train $L^2$-error (\%)}}
    & \multicolumn{3}{c}{\textbf{Test $L^2$-error (\%)}}
    \\ \cline{2-16}
    & 
    \multicolumn{1}{c}{MHNO} & \multicolumn{1}{c}{FNO-3d} & \multicolumn{1}{c|}{FNO-2d} & 
    \multicolumn{1}{c}{MHNO} & \multicolumn{1}{c}{FNO-3d} & \multicolumn{1}{c|}{FNO-2d} & 
    \multicolumn{1}{c}{MHNO} & \multicolumn{1}{c}{FNO-3d} & \multicolumn{1}{c|}{FNO-2d} & 
    \multicolumn{1}{c}{MHNO} & \multicolumn{1}{c}{FNO-3d} & \multicolumn{1}{c|}{FNO-2d} & 
    \multicolumn{1}{c}{MHNO} & \multicolumn{1}{c}{FNO-3d} & \multicolumn{1}{c}{FNO-2d} \\ \hline
    I   & 3,580,723 & 6,296,721 & 6,314,945 
        & \multicolumn{1}{c}{8.61}      & \multicolumn{1}{c}{42.42}     & \multicolumn{1}{c|}{41.92} 
        & \multicolumn{1}{c}{0.0597}    & \multicolumn{1}{c}{0.0530}    & \multicolumn{1}{c|}{0.2893} 
        & $0.84 \pm 1.57$               & $1.09 \pm 1.45$               & $0.33 \pm 0.22$
        & $2.64 \pm 3.64$               & $6.17 \pm 4.98$               & $6.81 \pm 4.76$ \\ 
    II  & 3,660,313 & 6,296,721 & 6,311,681 
        & \multicolumn{1}{c}{25.57}     & \multicolumn{1}{c}{112.66}    & \multicolumn{1}{c|}{135.35} 
        & \multicolumn{1}{c}{0.0379}    & \multicolumn{1}{c}{0.0235}    & \multicolumn{1}{c|}{0.2844} 
        & $0.81 \pm 0.15$                 & $1.03 \pm 0.23$             & $1.18 \pm 0.07$
        & $8.06 \pm 6.58$               & $8.44 \pm 6.56$               & $8,00 \pm 5,47$ \\ 
    III & 3,762,643 & 6,296,721 & 6,311,681 
        & \multicolumn{1}{c}{18.04}     & \multicolumn{1}{c}{100.55}    & \multicolumn{1}{c|}{182.63} 
        & \multicolumn{1}{c}{0.0347}    & \multicolumn{1}{c}{0.0173}    & \multicolumn{1}{c|}{0.2859} 
        & $0.89 \pm 0.21$                 & $0.97 \pm 0.15$             & $90.30 \pm 1.11$
        & $\textbf{2.55} \pm \textbf{3.13}$     & $2.61 \pm 3.38$       & $90.31 \pm 1.13$\\
    \bottomrule
    \end{tabular}%
    }
\end{table}

\begin{figure}[H]
    \centering
    \includegraphics[width=1.0\textwidth]{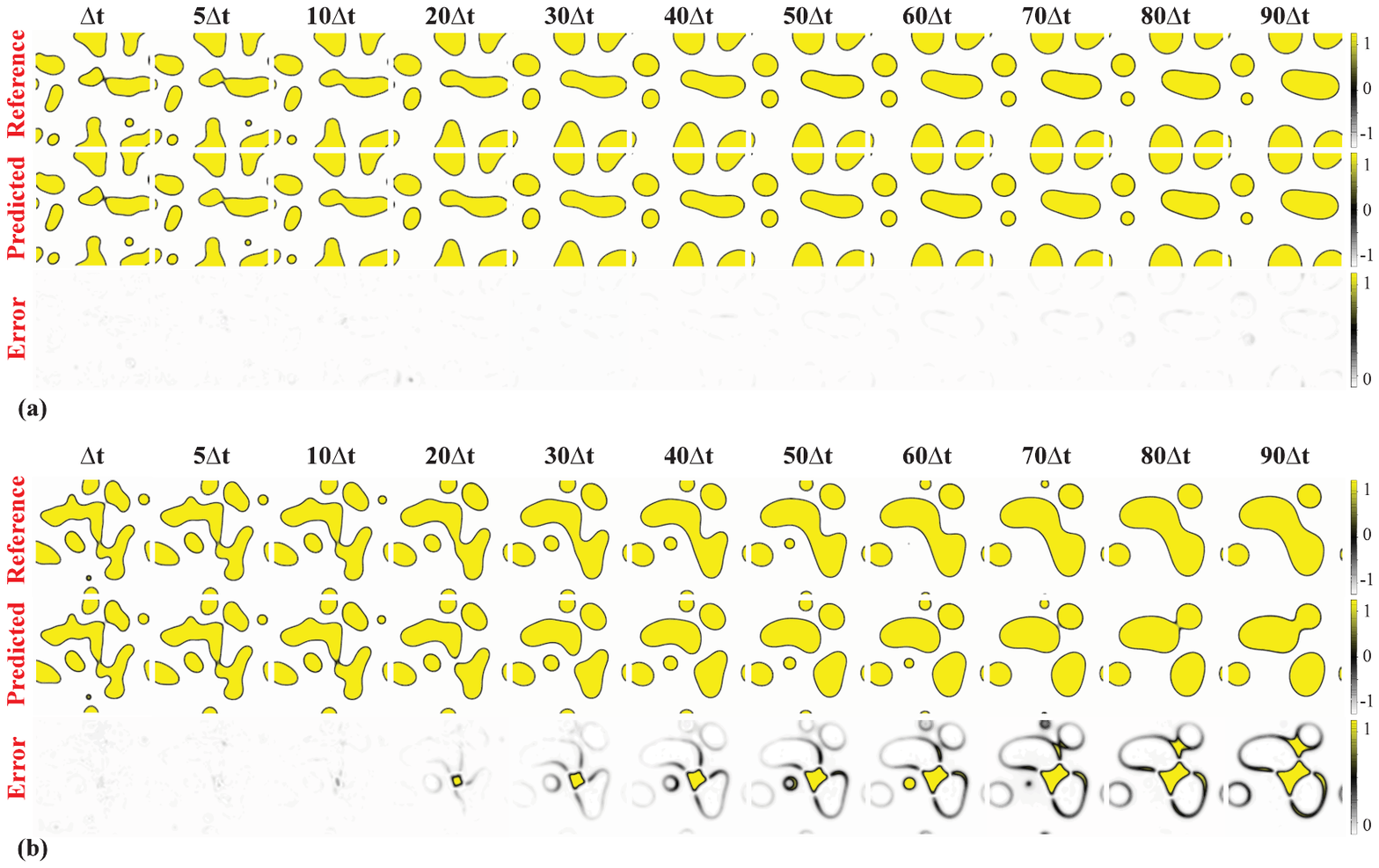}
    \caption{\textbf{Phase Field Predictions and Errors for Cahn-Hilliard Equation.} Visualization of the reference, predicted, and error fields for the Cahn-Hilliard equation over time steps (\(\Delta t, 5\Delta t, 10\Delta t, \ldots, 90\Delta t\)). (a) Results for a sample with an error close to the \textbf{mean} \(L^2\) error of the test dataset, showing accurate predictions and minimal discrepancies between the reference and predicted fields. (b) Results for the sample with the \textbf{highest} \(L^2\) error, highlighting regions where the model struggles to accurately capture the phase field dynamics, especially at later time steps. Error plots (bottom rows) highlight the regions of deviation.}
    \label{fig:CH2D}
\end{figure}

\begin{figure}[H]
    \centering
    \includegraphics[width=1.0\textwidth]{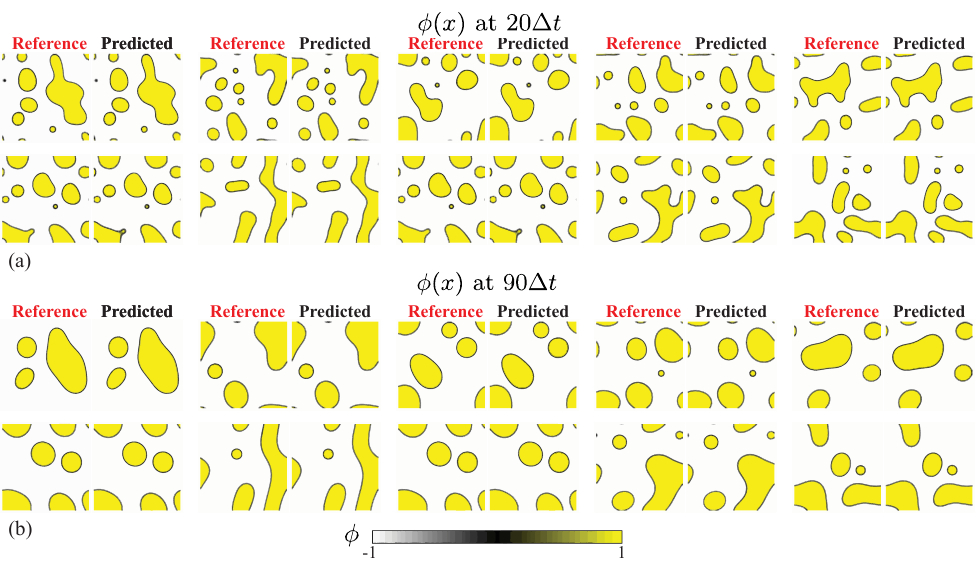}
    \caption{The Reference and MHNO predictions for 10 random samples from the dataset for spinodal decomposition.}
    \label{fig: CH2D-Comparison}
\end{figure}

Similar to the previous example, as shown in table \ref{tab:CH2D}, MHNO shows the lowest error in the third approach with many fewer parameters. In addition, figure \ref{fig:CH2D} illustrates the performance of phase field predictions for the 2D Cahn-Hilliard equation over various time steps, ranging from \(\Delta t\) to \(90\Delta t\). Two cases are shown: one with an error close to the mean \(L^2\) error of the test data set, and another with the highest \(L^2\) error. In both cases, the reference, predicted, and error fields are displayed to evaluate the accuracy of the model. In the first case (a), the predictions closely match the reference fields, with minimal discrepancies highlighted in the error plots, indicating that the model accurately captures the phase dynamics. In panel (b), the results correspond to the sample with the highest \(L^2\) error, revealing the difficulties the model encounters in capturing the phase field dynamics. As time progresses, especially for larger \(t\), there are noticeable differences between the predicted and reference fields, as highlighted in the error plots. This sample represents an outlier with a significantly higher error compared to most of the test cases, highlighting the variability in the model's prediction performance across different samples. Some random results for 10 sample trajectories from test datasets are shown in Fig. \ref{fig: CH2D-Comparison}

\subsection{Pattern Formation - Swift–Hohenberg Equation}

Pattern formation near bifurcation points arises in a wide range of physical, chemical, and biological systems. These phenomena are characterized by the spontaneous emergence of spatially periodic structures due to the interplay between diffusion, reaction, and nonlocal interactions. A classical example is Rayleigh–Bénard convection \cite{pandey2018turbulent}, where convective rolls form as a fluid layer is heated from below. Similarly, Turing patterns \cite{kuptsov2012hyperbolic} in reaction–diffusion systems illustrate how chemical concentrations self-organize into stable, often periodic configurations.

To model and study such behavior, researchers often employ PDEs that capture the essential dynamics near instability thresholds. One prototypical model for this is the Swift–Hohenberg equation \cite{swift1977hydrodynamic}, which captures the emergence and saturation of spatial patterns through a combination of linear instability and nonlinear saturation. It serves as a representative testbed for understanding how small perturbations in a homogeneous system evolve into regular, long-lived patterns. The Swift–Hohenberg equation can be written as \cite{swift1977hydrodynamic}:  
\begin{equation}
\frac{\partial \phi(x, t)}{\partial t} = - \phi^3(x, t) - (1-\epsilon) \phi(x,t) -2 \Delta \phi(x,t) - \Delta ^2 \phi(x, t) , \quad x \in [-1, 1]^2, \; t \in (0, T],
\end{equation}
\begin{equation}
\phi(x, 0) = \phi_0(x), \quad x \in [-1, 1]^2,
\end{equation}

where \( \phi(x, t) \) is the order parameter representing the state of the system and \( \epsilon \) is a real positive constant with respect to temperature. The term \(  \Delta ^2 \phi(x, t) \) stabilizes certain spatial wavelengths, leading to the formation of periodic structures. The Swift–Hohenberg equation introduces higher-order derivatives, contributing to its numerical solution's complexity. These derivatives control the length scales of the patterns, ensuring the stability of the periodic structures that emerge. This makes the equation an excellent candidate for testing neural operator-based solvers, as the dynamics involve both local and nonlocal interactions with strong spatial dependencies. We use \( \epsilon = 0.5 \), \( \Delta t_s = 0.5 \) and a resolution of \( 64 \times 64 \) for training and testing, simulating the dynamics over \( N_t = 90 \) time steps with \( \Delta t = 5 \). 

\begin{table}[H]
    \centering
    \caption{Performance comparison of neural operator models for solving the pattern formation problem across different approaches.}
    \label{tab: SH2D}
    \resizebox{\textwidth}{!}{%
    \begin{tabular}{l|lll|lll|lll|lll|lll}
    \toprule
    \multirow{2}{*}{\begin{tabular}[c]{@{}l@{}}\textbf{Temp.}\\ \textbf{Appr.}\end{tabular}}
    & \multicolumn{3}{c|}{\textbf{Parameters}}
    & \multicolumn{3}{c|}{\textbf{Training Time (s/epoch)}}
    & \multicolumn{3}{c|}{\textbf{Inference Time (s)}}
    & \multicolumn{3}{c|}{\textbf{Train $L^2$-error (\%)}}
    & \multicolumn{3}{c}{\textbf{Test $L^2$-error (\%)}}
    \\ \cline{2-16}
    & 
    \multicolumn{1}{c}{MHNO} & \multicolumn{1}{c}{FNO-3d} & \multicolumn{1}{c|}{FNO-2d} & 
    \multicolumn{1}{c}{MHNO} & \multicolumn{1}{c}{FNO-3d} & \multicolumn{1}{c|}{FNO-2d} & 
    \multicolumn{1}{c}{MHNO} & \multicolumn{1}{c}{FNO-3d} & \multicolumn{1}{c|}{FNO-2d} & 
    \multicolumn{1}{c}{MHNO} & \multicolumn{1}{c}{FNO-3d} & \multicolumn{1}{c|}{FNO-2d} & 
    \multicolumn{1}{c}{MHNO} & \multicolumn{1}{c}{FNO-3d} & \multicolumn{1}{c}{FNO-2d} \\ \hline
    I   & 3,240,291 & 6,296,721 & 3,225,153 
        & \multicolumn{1}{c}{10.71}         & \multicolumn{1}{c}{67.53}     & \multicolumn{1}{c|}{40.04} 
        & \multicolumn{1}{c}{0.0570}        & \multicolumn{1}{c}{0.0430}    & \multicolumn{1}{c|}{0.4237} 
        & $0.53 \pm 0.11$                   & $5.17 \pm 0.77$               & $0.89 \pm 0.33$
        & $\textbf{1.43} \pm \textbf{0.25}$ & $39.15 \pm 5.02$              & $3.04 \pm 1.86$ \\ 
    II  & 3,299,161 & 6,296,721 & 3,225,153 
        & \multicolumn{1}{c}{25.11}         & \multicolumn{1}{c}{107.86}    & \multicolumn{1}{c|}{130.75} 
        & \multicolumn{1}{c}{0.0442}        & \multicolumn{1}{c}{0.0214}    & \multicolumn{1}{c|}{0.2061} 
        & $0.86 \pm 0.07$                   & $6.99 \pm 0.61$               & $0.88 \pm 0.07$
        & $3.69 \pm 0.95$                   & $39.38 \pm 5.32$              & $3.62 \pm 1.47$ \\ 
    III & 3,374,851 & 6,296,721 & 3,225,153 
        & \multicolumn{1}{c}{25.15}         & \multicolumn{1}{c}{100.70}    & \multicolumn{1}{c|}{125.83} 
        & \multicolumn{1}{c}{0.0435}        & \multicolumn{1}{c}{0.0191}    & \multicolumn{1}{c|}{0.2116} 
        & $1.51 \pm 0.21$                   & $6.23 \pm 0.75$               & $3.23 \pm 0.36$
        & $2.66 \pm 0.80$                   & $15.14 \pm 3.59$              & $3.67 \pm 0.61$\\
    \bottomrule
    \end{tabular}%
    }
\end{table}

Table \ref{tab: SH2D} compares the performance of three neural operator models in solving the 2D Swift–Hohenberg equation under different approaches. MHNO exhibits the smallest error in the first approach and fastest training times across all approaches, demonstrating high computational efficiency. Additionally, it achieves the lowest deviation in \(L^2\)-error, indicating robust and reliable accuracy. In contrast to previous benchmark problems, Approach 1 outperforms the other two strategies for the Swift–Hohenberg equation. This may be attributed to the equation’s strong spatial coupling and nonlocal interactions, which benefit from training on dense, overlapping subsequences. By exposing the model to a wide range of short-term temporal patterns during training, Approach 1 enables better generalization in capturing the complex pattern formation dynamics characteristic of the Swift–Hohenberg system.

\begin{figure}[H]
    \centering
    \includegraphics[width=1.0\textwidth]{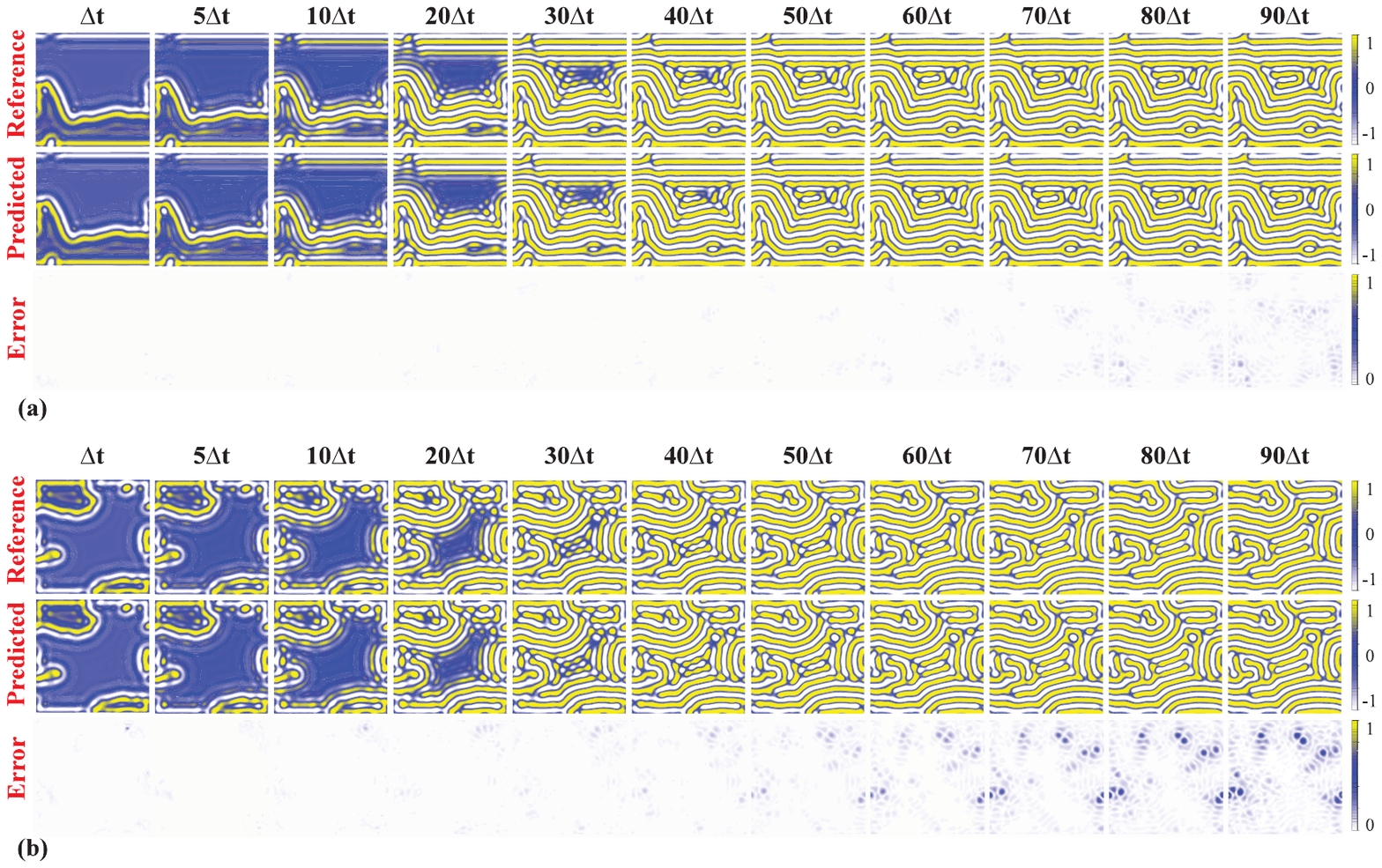}
    \caption{\textbf{Phase Field Predictions and Errors for Swift–Hohenberg Equation.} Visualization of the reference, predicted, and error fields for the Swift–Hohenberg equation over time steps (\(\Delta t, 5\Delta t, 10\Delta t, \ldots, 90\Delta t\)). (a) Results for a sample with an error close to the \textbf{mean} \(L^2\) error of the test dataset, showing accurate predictions and minimal discrepancies between the reference and predicted fields. (b) Results for the sample with the \textbf{highest} \(L^2\) error, highlighting regions where the model struggles to accurately capture the phase field dynamics, especially at later time steps. Error plots (bottom rows) highlight the regions of deviation.}
    \label{fig: SH2D}
\end{figure}

Figure \ref{fig: SH2D} displays the reference, predicted, and error fields for the 2D Swift–Hohenberg equation across various time steps. Panel (a) shows the predictions for a test sample with an \(L^2\) error close to the mean of the test dataset. Panel (b) highlights the results for the test sample with the highest \(L^2\) error, showing the model’s performance in accurately resolving certain phase field dynamics. Negligible differences between the reference and predicted fields emerge at later steps, with the error fields showing high performance of MHNO in modeling long-term dynamics in the Swift–Hohenberg equation.

\begin{figure}[H]
    \centering
    \includegraphics[width=\textwidth]{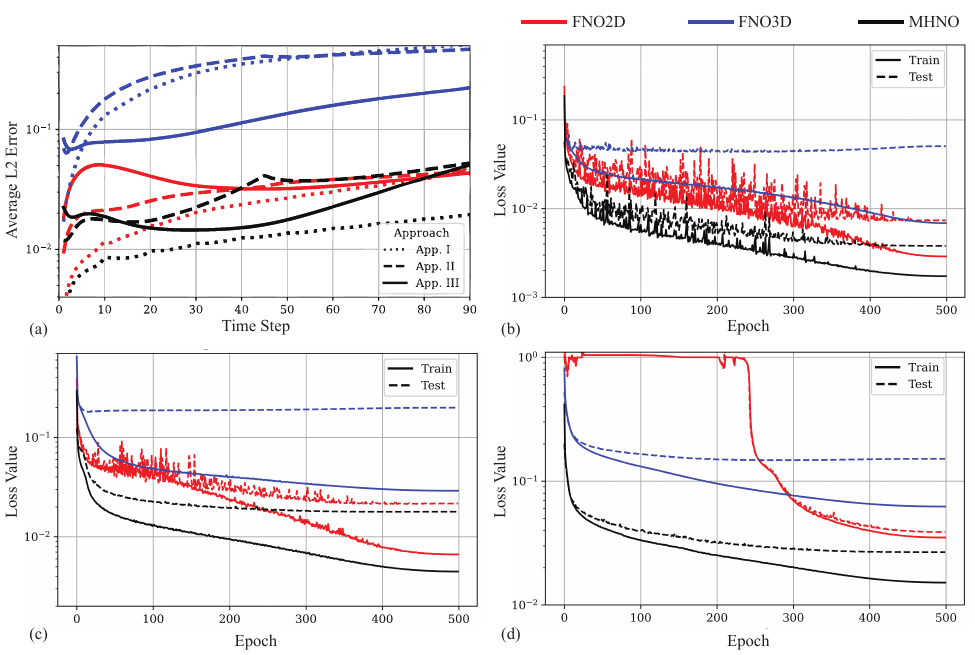}
    \caption{\textbf{Performance comparison of FNO2D, FNO3D, and MHNO on the pattern formation problem}. (a) Average $L^2$ error over time steps, computed across test samples. MHNO demonstrates the lowest error throughout the trajectory. (b–d) Training loss convergence for Approaches I (windowed subsequence), II (two-window), and III (full-trajectory), respectively. MHNO consistently converges faster and to a lower loss across all training strategies.}
    \label{fig: SH2D-Comparison}
\end{figure}

Figure \ref{fig: SH2D-Comparison} presents the evaluation of different neural operator models on the pattern formation problem across three training approaches. Subfigure \ref{fig: SH2D-Comparison}(a) shows the average $L^2$ error over time steps, averaged across test samples. MHNO consistently achieves the lowest error throughout the temporal horizon, demonstrating superior long-term accuracy and stability compared to FNO2D and FNO3D. Notably, FNO2D outperforms FNO3D, indicating that capturing spatiotemporal features jointly (as in 3D convolutions) is less effective than treating spatial and temporal components separately in the Swift-Hohenberg equation. Subfigures \ref{fig: SH2D-Comparison} (b), (c), and (d) depict the loss convergence curves for training Approaches I (Windowed Subsequence), II (Two-Window), and III (Full-Trajectory), respectively. In all three settings, MHNO converges faster and to a lower loss value than the other methods, confirming its training efficiency and robustness across various supervision schemes. The convergence trend is especially pronounced in Approach I, which aligns with MHNO's strong performance in prediction accuracy under this training regime. This suggests that MHNO effectively exploits overlapping temporal context, which is particularly beneficial for modeling the nonlocal, pattern-forming dynamics of the Swift–Hohenberg system.

\subsection{Atomic-Scale Modeling - Phase Field Crystal Equation}

Atomic-scale processes such as crystal nucleation \cite{granasy2002nucleation}, material growth \cite{elder2007phase}, and defect dynamics \cite{skogvoll2022phase} play a critical role in determining the structure and properties of materials. Modeling these phenomena requires capturing both the long-term dynamics of material evolution and the intrinsic spatial periodicity at the atomic scale. One effective approach is to use phase field methods that incorporate such fine-scale features directly into the model formulation.

To this end, we have considered a phase field framework capable of resolving atomic-scale structure, the Phase Field Crystal (PFC) model, which is a sixth-order PDE. The PFC equation captures the evolution of a phase field variable \( \phi(x, t) \) that represents the atomic density deviation from its average value. Unlike the previous phase field models, the PFC equation naturally incorporates atomic-scale spatial periodicity, making it a powerful tool for studying atomic-scale phenomena. The governing equation is given by \cite{elder2002modeling}:
\begin{equation}
\frac{\partial \phi(x, t)}{\partial t} = \Delta \left[ \phi^3 + (1 - \epsilon) \phi + 2\Delta \phi + \Delta^2 \phi \right], \quad x \in [-1, 1]^2, \; t \geq 0,
\end{equation}
\begin{equation}
\phi(x, 0) = \phi_0(x), \quad x \in [-1, 1]^2,
\end{equation}
where \( \phi(x, t) \) is the phase field variable, \( \epsilon \) is a parameter controlling the degree of undercooling, and \( \Delta \) is the Laplace operator. The terms \( \phi^3 \) and \( (1 - \epsilon)\phi \) govern the local dynamics, while \( 2\Delta \phi + \Delta^2 \phi \) stabilize the periodic structure characteristic of crystalline phases. The Laplacian \( \Delta \) in the evolution equation enforces smooth transitions between atomic layers. The PFC equation's higher-order derivatives (up to sixth-order) and the need to resolve fine-scale periodic features make it computationally demanding, particularly for simulating large systems over long timescales. It serves as an excellent benchmark problem for evaluating the capabilities of neural operator-based solvers in handling atomic-scale dynamics. For the experiments, we use a \(\epsilon=0.025\), $\Delta t_s = 0.1$, fixed domain resolution of \( 64 \times 64 \), with a simulation duration of \( N_t = 100 \) time steps and \( \Delta t = 10 \). 

\begin{table}[H]
    \centering
    \caption{Performance comparison of neural operator models for atomic scale modeling across different approaches}
    \label{tab: PFC2D}
    \resizebox{\textwidth}{!}{%
    \begin{tabular}{l|lll|lll|lll|lll|lll}
    \toprule
    \multirow{2}{*}{\begin{tabular}[c]{@{}l@{}}\textbf{Temp.}\\ \textbf{Appr.}\end{tabular}}
    & \multicolumn{3}{c|}{\textbf{Parameters}}
    & \multicolumn{3}{c|}{\textbf{Training Time (s/epoch)}}
    & \multicolumn{3}{c|}{\textbf{Inference Time (s)}}
    & \multicolumn{3}{c|}{\textbf{Train $L^2$-error (\%)}}
    & \multicolumn{3}{c}{\textbf{Test $L^2$-error (\%)}}
    \\ \cline{2-16}
    & 
    \multicolumn{1}{c}{MHNO} & \multicolumn{1}{c}{FNO-3d} & \multicolumn{1}{c|}{FNO-2d} & 
    \multicolumn{1}{c}{MHNO} & \multicolumn{1}{c}{FNO-3d} & \multicolumn{1}{c|}{FNO-2d} & 
    \multicolumn{1}{c}{MHNO} & \multicolumn{1}{c}{FNO-3d} & \multicolumn{1}{c|}{FNO-2d} & 
    \multicolumn{1}{c}{MHNO} & \multicolumn{1}{c}{FNO-3d} & \multicolumn{1}{c|}{FNO-2d} & 
    \multicolumn{1}{c}{MHNO} & \multicolumn{1}{c}{FNO-3d} & \multicolumn{1}{c}{FNO-2d} \\ \hline
    I   & 2,388,323                         & 4,197,937                     & 2,984,429 
        & \multicolumn{1}{c}{11.95}         & \multicolumn{1}{c}{32.64}     & \multicolumn{1}{c|}{60.22} 
        & \multicolumn{1}{c}{0.0586}        & \multicolumn{1}{c}{0.0478}    & \multicolumn{1}{c|}{0.3356} 
        & $0.51 \pm 0.78$                   & $4.86 \pm 1.68$               & $2.77 \pm 1.41$
        & $\textbf{0.77} \pm \textbf{0.50}$ & $11.66 \pm 6.03$              & $7.06 \pm 15.08$ \\ 
    II  & 2,447,193                         & 6,296,721                     & 4,209,281 
        & \multicolumn{1}{c}{32.75}         & \multicolumn{1}{c}{113.17}    & \multicolumn{1}{c|}{145.77} 
        & \multicolumn{1}{c}{0.0457}        & \multicolumn{1}{c}{0.0228}    & \multicolumn{1}{c|}{0.2096} 
        & $0.71 \pm 0.25$                   & $2.93 \pm 0.29$               & $2.74 \pm 1.11$
        & $3.96 \pm 3.03$                   & $20.58 \pm 7.91$              & $10.86 \pm 6.23$ \\ 
    III & 2,522,883                         & 6,296,721 & 4,209,281
        & \multicolumn{1}{c}{26.45}         & \multicolumn{1}{c}{101.13}    & \multicolumn{1}{c|}{141.34} 
        & \multicolumn{1}{c}{0.0410}        & \multicolumn{1}{c}{0.0179}    & \multicolumn{1}{c|}{0.2356} 
        & $0.99 \pm 0.31$                     & $2.01 \pm 0.45$             & $65.93 \pm 5.47$
        & $1.76 \pm 1.13$                   & $3.71 \pm 2.14$               & $66.02 \pm 5.42$\\
    \bottomrule
    \end{tabular}%
    }
\end{table}

Table \ref{tab: PFC2D} highlights the performance of neural operator models in solving the Phase Field Crystal equation under varying approaches. The evaluation criteria include the number of parameters, the training time per epoch, inference time, and the \(L^2\)-error. MHNO consistently demonstrates the fewest parameters and the shortest training duration per epoch, indicating it is more computationally efficient than both FNO variants. Additionally, MHNO delivers the most accurate predictions, with the smallest \(L^2\)-error and error deviation. Similar to the Swift-Hohenberg equation, approach 1 outperforms the other two strategies. 

\begin{figure}[H]
    \centering
    \includegraphics[width=1.0\textwidth]{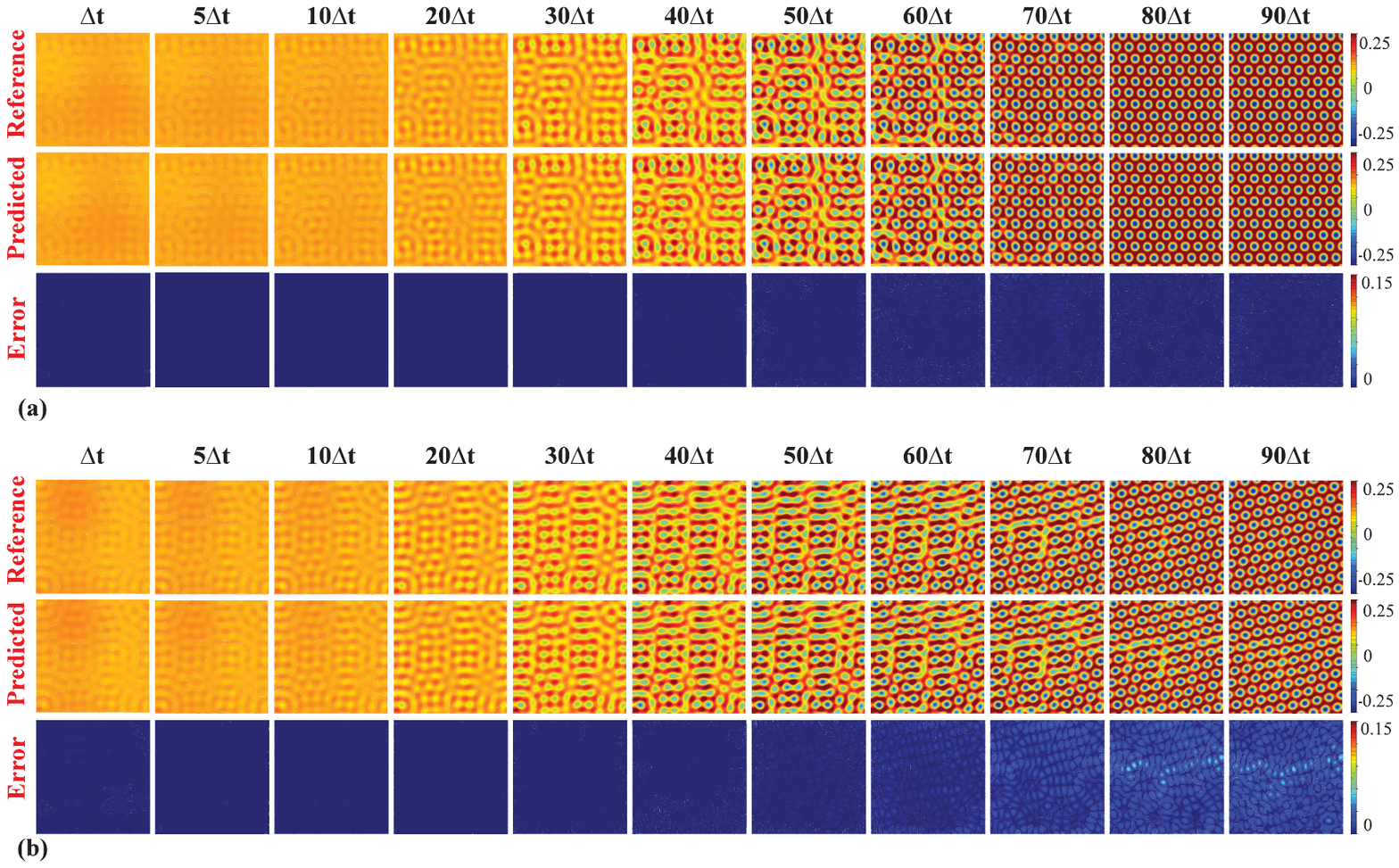}
    \caption{\textbf{Phase Field Predictions and Errors for 2D Phase Field Crystal Equation.} Visualization of the reference, predicted, and error fields for the Phase Field Crystal equation over time steps (\(\Delta t, 5\Delta t, 10\Delta t, \ldots, 90\Delta t\)). (a) Results for a sample with an error close to the \textbf{mean} \(L^2\) error of the test dataset, showing accurate predictions and minimal discrepancies between the reference and predicted fields. (b) Results for the sample with the \textbf{highest} \(L^2\) error, highlighting regions where the model struggles to accurately capture the phase field dynamics, especially at later time steps. Error plots (bottom rows) highlight the regions of deviation.}
    \label{fig: PFC2D}
\end{figure}

\begin{figure}[H]
    \centering
    \includegraphics[width=\textwidth]{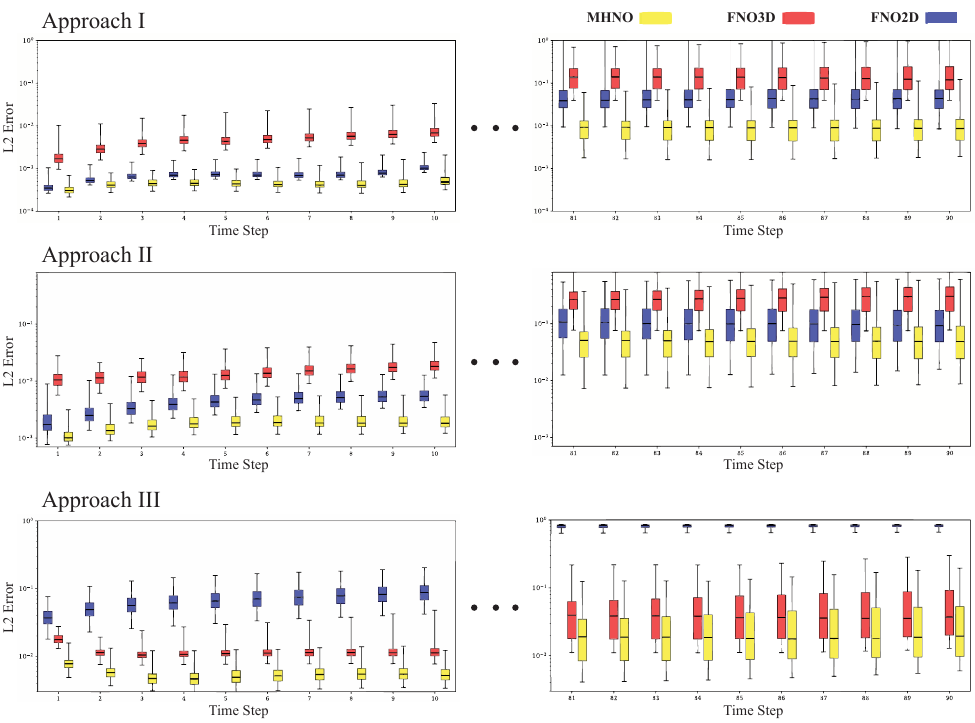}
    \caption{Box plots of $L^2$ error over the first ten and last ten time steps for the PFC equation, evaluated across different test samples. Each row corresponds to one of the three training approaches, with comparisons between FNO2D, FNO3D, and MHNO. All methods show some increase in error at later time steps, but MHNO maintains consistently low error and variability, indicating strong long-term predictive accuracy.}
    \label{fig: PFC-Comparison}
\end{figure}

Figure \ref{fig: PFC2D} illustrates the reference, predicted, and error fields for the Phase Field Crystal equation across multiple time steps. Panel (a) displays predictions for a representative test sample with an \(L^2\)-error close to the dataset’s mean error. The predicted fields closely align with the reference across all time steps, with minimal differences visible in the error fields (bottom row). In addition, panel (b) presents results for the test sample with the largest \(L^2\)-error, highlighting scenarios where the model encounters challenges. As time progresses, deviations between the reference and predicted fields become more pronounced, particularly at later time steps. Moreover, Fig. \ref{fig: PFC-Comparison} presents box plots of the $L^2$ error across different test samples for the first ten and last ten time steps of the PFC simulation. Each row corresponds to one of the three training approaches. Across all approaches, a general increase in error is observed at later time steps, reflecting the growing challenge of long-term prediction in the PFC dynamics. However, MHNO consistently maintains lower median error and narrower variability, demonstrating strong temporal generalization and robustness, especially in the later stages of evolution where predictive degradation is most likely. In contrast, FNO exhibits substantial error growth at later steps, particularly under Approach III, where it fails to maintain predictive accuracy. 

\subsection{Molecular Beam Epitaxy (MBE) Growth Model}

The Molecular Beam Epitaxy (MBE) growth model is a PDE used to describe the process of depositing a thin single-crystal layer on a substrate using molecular beams. This technique, widely employed in material science and nanotechnology, enables precise control over layer thickness and composition, making it essential for fabricating advanced semiconductor devices \cite{eaglesham1995semiconductor} and nanostructures \cite{hasan2023selective}. The MBE model captures the morphological evolution of the thin film surface during growth, taking into account surface diffusion and conservation of mass.

The governing equation for the phase field variable \( \phi(x, t) \), which represents the surface height, is given by \cite{qiao2011adaptive, chen2018regularized}:  
\begin{equation}
\frac{\partial \phi(x, t)}{\partial t} = -\epsilon \Delta^2 \phi - \nabla \cdot \left[ (|1 - \nabla \phi|^2) \nabla \phi \right], \quad x \in [-\pi, \pi], \; t \geq 0,
\end{equation}
\begin{equation}
\phi(x, 0) = \phi_0(x), \quad x \in [-\pi, \pi],
\end{equation}
where \( \epsilon > 0 \) is a parameter controlling the influence of the surface diffusion term \( -\epsilon \Delta^2 \phi \), which drives the system toward a smoother surface morphology. The nonlinear term \( \nabla \cdot \left[ (|1 - \nabla \phi|^2) \nabla \phi \right] \) accounts for the impact of local surface gradients on the growth process. This term introduces anisotropic effects that can lead to the formation of complex surface structures, such as mounds or ripples. The MBE growth model poses significant computational challenges due to the fourth-order derivatives and the nonlinear gradient term, both of which are sensitive to the spatial resolution. These features make the equation a valuable benchmark for testing neural operator-based solvers in simulating surface dynamics. For the experiments, the dynamics are simulated with \( \Delta t_s = 0.05 \) on a domain with a resolution of \( 64 \times 64 \), using a total of \( N_t = 100 \) time steps and \( \Delta t = 5 \). 

\begin{table}[H]
    \centering
    \caption{Performance comparison of neural operator models for MBE modelling across different approaches.}
    \label{tab:MBE}
    \resizebox{\textwidth}{!}{%
    \begin{tabular}{l|lll|lll|lll|lll|lll}
    \toprule
    \multirow{2}{*}{\begin{tabular}[c]{@{}l@{}}\textbf{Temp.}\\ \textbf{Appr.}\end{tabular}}
    & \multicolumn{3}{c|}{\textbf{Parameters}}
    & \multicolumn{3}{c|}{\textbf{Training Time (s/epoch)}}
    & \multicolumn{3}{c|}{\textbf{Inference Time (s)}}
    & \multicolumn{3}{c|}{\textbf{Train $L^2$-error (\%)}}
    & \multicolumn{3}{c}{\textbf{Test $L^2$-error (\%)}}
    \\ \cline{2-16}
    & 
    \multicolumn{1}{c}{MHNO} & \multicolumn{1}{c}{FNO-3d} & \multicolumn{1}{c|}{FNO-2d} & 
    \multicolumn{1}{c}{MHNO} & \multicolumn{1}{c}{FNO-3d} & \multicolumn{1}{c|}{FNO-2d} & 
    \multicolumn{1}{c}{MHNO} & \multicolumn{1}{c}{FNO-3d} & \multicolumn{1}{c|}{FNO-2d} & 
    \multicolumn{1}{c}{MHNO} & \multicolumn{1}{c}{FNO-3d} & \multicolumn{1}{c|}{FNO-2d} & 
    \multicolumn{1}{c}{MHNO} & \multicolumn{1}{c}{FNO-3d} & \multicolumn{1}{c}{FNO-2d} \\ \hline
    I   & 2,402,867                             & 4,197,937 & 3,225,153
        & \multicolumn{1}{c}{10.93}             & \multicolumn{1}{c}{73.84}         & \multicolumn{1}{c|}{42.14} 
        & \multicolumn{1}{c}{0.0578}            & \multicolumn{1}{c}{0.0188}        & \multicolumn{1}{c|}{0.2362} 
        & $0.37 \pm 0.10$                       & $0.11 \pm 0.36$                   & $0.92 \pm 0.14$
        & $10.40 \pm 7.19$                      & $11.08 \pm 7.80$                  & $29.83 \pm 12.47$ \\ 
    II  & 2,518,297                             & 4,197,937 & 3,225,153
        & \multicolumn{1}{c}{31.89}             & \multicolumn{1}{c}{85.60}         & \multicolumn{1}{c|}{129.75}
        & \multicolumn{1}{c}{0.0442}            & \multicolumn{1}{c}{0.0178}        & \multicolumn{1}{c|}{0.2040}
        & $0.59 \pm 0.09$                       & $0.47 \pm 0.08$                   & $1.30 \pm 0.18$
        & $7.69 \pm 3.34$                       & $7.29 \pm 2.92$                   & $17.05 \pm 7.29$ \\ 
    III & 2,666,707                             & 4,197,937                         & 3,225,153 
        & \multicolumn{1}{c}{26.51}             & \multicolumn{1}{c}{63.70}         & \multicolumn{1}{c|}{126.55} 
        & \multicolumn{1}{c}{0.0390}            & \multicolumn{1}{c}{0.0471}        & \multicolumn{1}{c|}{0.2079} 
        & $0.96 \pm 0.19$                       & $0.74 \pm 0.15$                   & $76.12 \pm 10.70$
        & $\textbf{1.67} \pm \textbf{0.69}$     & $1.73 \pm 0.78$                   & $77.25 \pm 11.02$ \\
    \bottomrule
    \end{tabular}%
    }
\end{table}

\begin{figure}[H]
    \centering
    \includegraphics[width=1.0\textwidth]{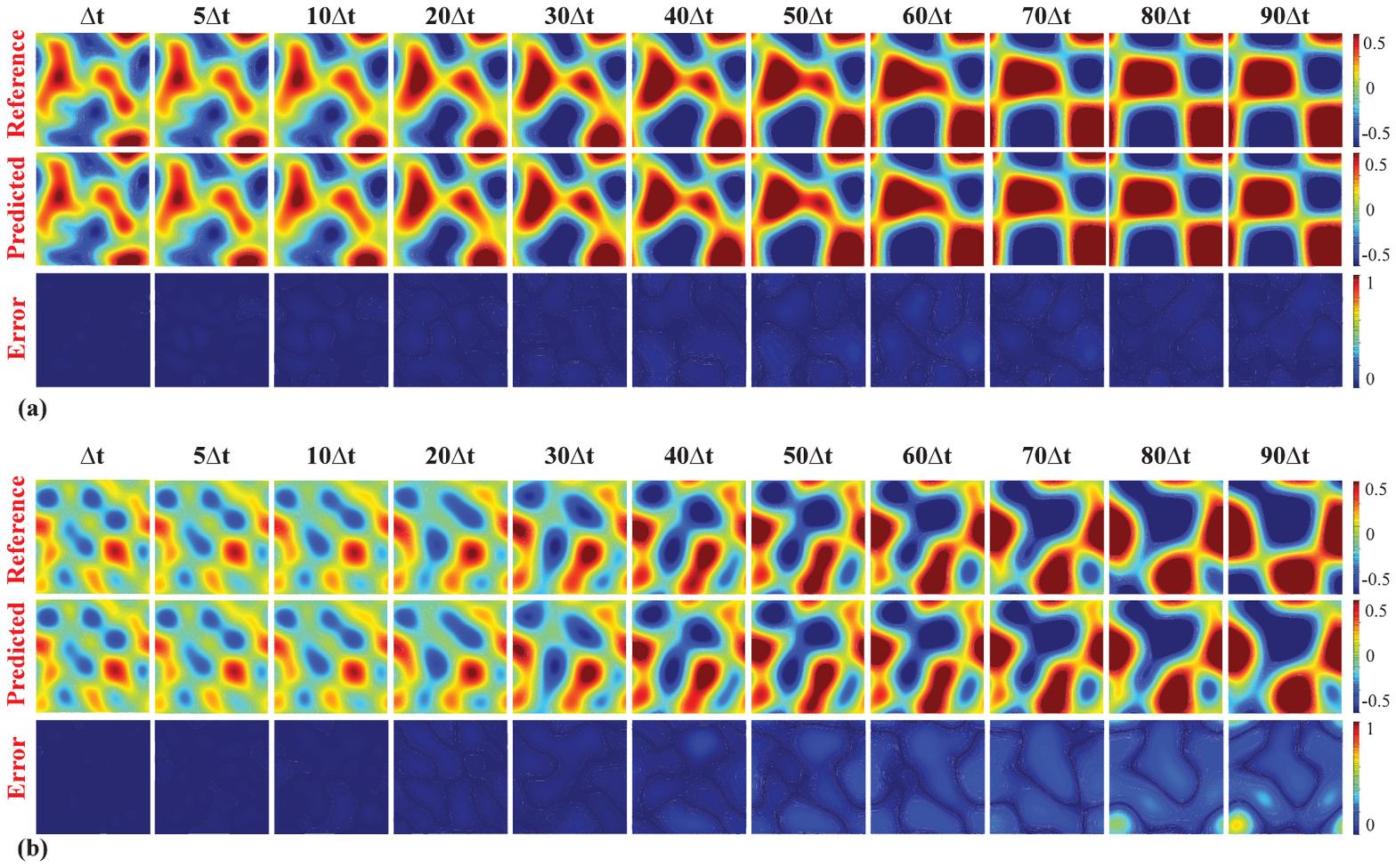}
    \caption{\textbf{Phase Field Predictions and Errors for 2D Molecular Beam Epitaxy Equation.} Visualization of the reference, predicted, and error fields for the Molecular Beam Epitaxy equation over time steps (\(\Delta t, 5\Delta t, 10\Delta t, \ldots, 90\Delta t\)). (a) Results for a sample with an error close to the \textbf{mean} \(L^2\) error of the test dataset, showing accurate predictions and minimal discrepancies between the reference and predicted fields. (b) Results for the sample with the \textbf{highest} \(L^2\) error, highlighting regions where the model struggles to accurately capture the phase field dynamics, especially at later time steps. The error plots (bottom rows) are scaled to \(1/4\) of the reference or predicted fields to better emphasize regions of deviation.}
    \label{fig:MBE2D}
\end{figure}

Table \ref{tab:MBE} presents a performance evaluation of the methods in addressing the Molecular Beam Epitaxy equation at various temporal approaches. The MHNO model demonstrates superior computational performance with the fewest parameters and shortest training times per epoch for all temporal resolutions. In terms of accuracy, MHNO achieves lower \(L^2\)-errors compared to FNO models. Moreover, figure \ref{fig:MBE2D} visualizes the model’s predictions, along with error fields, for the Molecular Beam Epitaxy equation at various time steps (\(\Delta t, 5\Delta t, 10\Delta t, \ldots, 100\Delta t\)). Panel (a) corresponds to a sample with a mean \(L^2\)-error close to the dataset average, revealing strong alignment between the predicted and reference fields. Panel (b) depicts a challenging test case, representing the sample with the largest \(L^2\)-error. Here, deviations between the reference and predicted fields emerge as time progresses, particularly at later stages. The corresponding error fields, scaled down to one-fourth of the predicted and reference field values, effectively illustrate regions of pronounced inaccuracies. 

\begin{figure}[H]
    \centering
    \includegraphics[width=\textwidth]{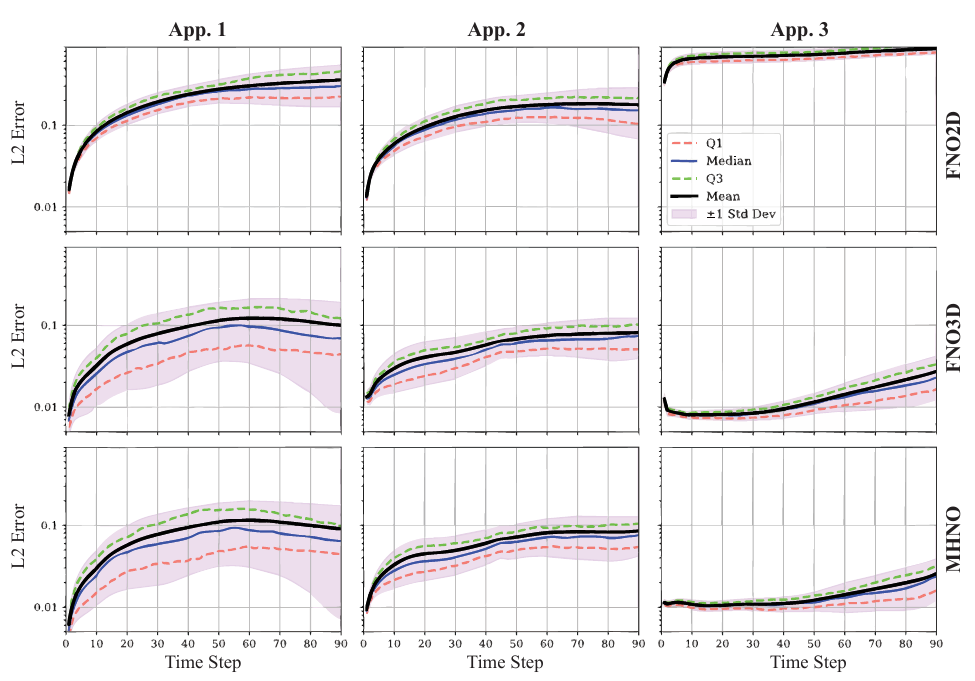}
    \caption{Comparison of L2 error statistics over time steps for different methods and different approaches}
    \label{fig: MBE-Comparision}
\end{figure}

Figure \ref{fig: MBE-Comparision} presents a comparison of L2 error statistics over time steps for the methods. In this figure, the L2 error is plotted against time steps for each method, providing insight into how the accuracy of each method evolves as the prediction horizon extends. Among the methods compared, MHNO consistently exhibits the lowest L2 error across the majority of time steps. This indicates that MHNO delivers the most accurate predictions over time compared to the other methods evaluated.

\section{Conclusion} \label{sec:Conclusion}

In this work, we introduced the Multi-Head Neural Operator (MHNO), a novel extension of neural operator frameworks designed to address the challenges of solving time-dependent phase field problems over long temporal horizons. By replacing the global projection operator with a series of time-specific neural networks and incorporating explicit connections between adjacent time steps, MHNO establishes a more effective mechanism for capturing temporal dependencies and long-term dynamics in solution fields.  

Through experiments on a diverse set of physical phenomenon which needs modelling interfacial dynamics, including the antiphase boundary motion, spinodal decomposition, pattern formation, atomic scale modeling, and molecular beam epitaxy growth model, we demonstrated that MHNO achieves superior accuracy, scalability, and computational efficiency compared to existing neural operator methods such as FNO-2d and FNO-3d. MHNO mitigates error accumulation through the prediction of all time steps in a single forward pass and avoids the parameter explosion associated with high-dimensional tensor representations in FNO-3d. Furthermore, the incorporation of temporal connections between time steps ensures that critical information is retained and utilized throughout the prediction process, improving its ability to model complex dynamic systems.  

The structural innovations in MHNO, inspired by message-passing principles, offer a new perspective on how neural operators can be extended to model time-dependent phenomena. By preserving the key strengths of neural operators, such as the use of Fourier integral operators for scalability, and addressing their limitations in temporal modeling, MHNO demonstrates strong potential for application in a wide range of scientific and engineering problems. Future work could explore further extensions to MHNO, including integration of physics into the networks, and applications to higher-dimensional systems and multi-phase phenomena. With its strong foundation and empirical performance, MHNO represents a significant step forward in the development of neural operator methodologies for time-dependent PDEs.

\section*{Declaration of Competing Interest}
The authors declare that they have no known competing financial interests or personal relationships that could have appeared to influence the work reported in this paper.

\section*{Acknowledgments}
The authors would like to acknowledge the support provided by the German Academic Exchange Service (DAAD) through a scholarship awarded to Mohammad Sadegh Eshaghi during this research.

\begin{appendices}

\section{Fourier-Spectral Method} \label{appendix:A}

In this section, we outline the Fourier-spectral method used for solving phase field models in two-dimensional space, \(\Omega = [l_x, r_x] \times [l_y, r_y]\). This method provides an stable approach to numerically solving these equations. Let \(N_x\) and \(N_y\) be positive even integers, and \(L_x = r_x - l_x\), \(L_y = r_y - l_y\) represent the lengths in the \(x\)- and \(y\)-directions, respectively. The spatial step sizes are given as \(h_x = L_x / N_x\) and \(h_y = L_y / N_y\). The discretized spatial points are \((x_m, y_n) = (l_x + m h_x, l_y + n h_y)\), where \(0 \leq m \leq N_x\) and \(0 \leq n \leq N_y\) are integers.

Let \(\phi_k^{m, n}\) be an approximation of \(\phi(x_m, y_n, t_k)\), where \(t_k = k\Delta t\) and \(\Delta t\) is the temporal step size. For the given data \(\{\phi_k^{m, n} \mid m = 1, \dots, N_x, n = 1, \dots, N_y\}\), the discrete Fourier transform (DFT) is defined as:  
\begin{equation}
\hat{\phi}_k^{p, q} = \sum_{m=1}^{N_x} \sum_{n=1}^{N_y} \phi_k^{m, n} e^{-i (\xi_p x_m + \eta_q y_n)}, \quad -\frac{N_x}{2} + 1 \leq p \leq \frac{N_x}{2}, \quad -\frac{N_y}{2} + 1 \leq q \leq \frac{N_y}{2},
\end{equation}

where \(\xi_p = \frac{2\pi p}{L_x}\) and \(\eta_q = \frac{2\pi q}{L_y}\).

The inverse discrete Fourier transform (IDFT) is expressed as:  
\begin{equation}
\phi_k^{m, n} = \frac{1}{N_x N_y} \sum_{p=-N_x/2+1}^{N_x/2} \sum_{q=-N_y/2+1}^{N_y/2} \hat{\phi}_k^{p, q} e^{i (\xi_p x_m + \eta_q y_n)}.
\end{equation}

Using the DFT, spectral derivatives can be calculated in Fourier space without directly differentiating in physical space. For a sufficiently smooth function \(\phi(x, y, t)\), the derivatives are obtained as:
\begin{equation}
\frac{\partial}{\partial x}\phi(x, y, t) = \frac{1}{N_x N_y} \sum_{p=-N_x/2+1}^{N_x/2} \sum_{q=-N_y/2+1}^{N_y/2} (i\xi_p) \hat{\phi}(\xi_p, \eta_q, t) e^{i(\xi_p x + \eta_q y)},
\end{equation}

\begin{equation}
\frac{\partial}{\partial y}\phi(x, y, t) = \frac{1}{N_x N_y} \sum_{p=-N_x/2+1}^{N_x/2} \sum_{q=-N_y/2+1}^{N_y/2} (i\eta_q) \hat{\phi}(\xi_p, \eta_q, t) e^{i(\xi_p x + \eta_q y)}.
\end{equation}

By applying these differentiation rules twice, the Laplacian can be represented in the Fourier space as:
\begin{equation}
\Delta\phi(x, y, t) = \frac{1}{N_x N_y} \sum_{p=-N_x/2+1}^{N_x/2} \sum_{q=-N_y/2+1}^{N_y/2} -(\xi_p^2 + \eta_q^2) \hat{\phi}(\xi_p, \eta_q, t) e^{i(\xi_p x + \eta_q y)},
\end{equation}

The following provides the solution of the Allen–Cahn equation as an illustrative example of the method. The discretized form is given by:
\begin{equation}
\frac{\phi_k^{m, n} - \phi_{k-1}^{m, n}}{\Delta t} = -\frac{2\phi_k^{m, n} + g(\phi_{k-1}^{m, n})}{\epsilon^2} + (\Delta \phi_k)^{m, n},
\end{equation}
where \(g(\phi) = \phi^3 - 3\phi\). Transitioning to the discrete Fourier space, the equation becomes:
\begin{equation}
\frac{\hat{\phi}_k^{p, q} - \hat{\phi}_{k-1}^{p, q}}{\Delta t} = -\frac{2\hat{\phi}_k^{p, q} + \hat{g}_{k-1}^{p, q}}{\epsilon^2} - (\xi_p^2 + \eta_q^2)\hat{\phi}_k^{p, q}.
\end{equation}

Rearranging for \(\hat{\phi}_k^{p, q}\), we derive:
\begin{equation}
\hat{\phi}_k^{p, q} = \frac{\epsilon^2 \hat{\phi}_{k-1}^{p, q} - \Delta t \hat{f}_{k-1}^{p, q}}{\epsilon^2 + \Delta t \left[2 + \epsilon^2 (\xi_p^2 + \eta_q^2)\right]}.
\end{equation}

The updated solution in physical space is then computed using the inverse Fourier transform:
\begin{equation}
\phi_k^{m, n} = \frac{1}{N_x N_y} \sum_{p=-N_x/2+1}^{N_x/2} \sum_{q=-N_y/2+1}^{N_y/2} \hat{\phi}_k^{p, q} e^{i(\xi_p x_m + \eta_q y_n)}.
\end{equation}

This completes the numerical solution of the Allen-Cahn equation using the Fourier-spectral method \cite{yoon2020fourier}. This method allows for efficient calculation of spatial derivatives and Laplacians in the Fourier space, ensuring high accuracy and stability. Additional details on implementing the Fourier-spectral method for various phase field models can be found in \cite{yoon2020fourier}.

\section{Hyperparameters} \label{appendix:B}

We carefully tuned the hyperparameters of all models to ensure a fair and meaningful comparison across different methods, problems, and solution approaches. This tuning process was carried out independently for each setting, taking into account the specific characteristics and requirements of the underlying equation and the model architecture. The results reported for both the baseline models and the proposed method correspond to the best performance achieved during our experiments, following extensive testing of various hyperparameter configurations. As an illustrative example, Table~\ref{tab:hyperparameters_tuning} presents the performance of the FNO2d model applied to the Allen–Cahn equation under Approach III. This table includes key hyperparameters such as the learning rate, number of epochs, feature dimensions, and network depths. For all other combinations of methods, equations, and approaches considered in this study, a similar hyperparameter tuning procedure was performed to ensure consistency and optimal performance across the board.

\begin{table}[H]
    \centering
    \caption{Hyperparameter configurations and corresponding $L^2$-Errors for FNO2d applied to the Allen–Cahn equation using Approach I.}
    \label{tab:hyperparameters_tuning}
    \resizebox{9.6cm}{!}{%
    \begin{tabular}{l|cccccccccc}
\multicolumn{1}{c|}{Run ID} & \textbf{\(Lr\)} & \textbf{Epoch}   & \textbf{\(M\)} & \textbf{\(W\)} & \textbf{\(W_\mathcal{Q}\)} & \textbf{\(N_l\)} & \textbf{\(N_\mathcal{Q}\)} & \textbf{\(L^2\)-Error} \\ \hline
1  & 0.00005 & 500 & 16 & 16 & 32 & 4 & 2 & 0.364747 \\ 
2  & 0.00010 & 500 & 16 & 16 & 32 & 4 & 2 & 0.145796 \\
3  & 0.00050 & 500 & 16 & 16 & 32 & 4 & 2 & 0.509595 \\
4  & \textbf{0.00100} & 500 & 16 & 16 & 32 & 4 & 2 & \textbf{0.118131} \\
5  & 0.00500 & 500 & 16 & 16 & 32 & 4 & 2 & 0.514202 \\ \hline
6  & 0.0001 & 500 & 6  & 16 & 128 & 4 & 4 & 0.544351 \\ 
7  & 0.0001 & 500 & 8  & 16 & 128 & 4 & 4 & 0.476792 \\
8  & 0.0001 & 500 & 10 & 16 & 128 & 4 & 4 & 0.374494 \\ 
9  & 0.0001 & 500 & 12 & 16 & 128 & 4 & 4 & 0.323541 \\
10 & 0.0001 & 500 & 14 & 16 & 128 & 4 & 4 & 0.308784 \\ 
11 & 0.0001 & 500 & 16 & 16 & 128 & 4 & 4 & 0.293158 \\
12 & 0.0001 & 500 & \textbf{18} & 16 & 128 & 4 & 4 & \textbf{0.288367} \\ \hline
13 & 0.0001 & 500 & 10 & 12 & 128 & 4 & 4 & 0.330834 \\ 
14 & 0.0001 & 500 & 10 & 16 & 128 & 4 & 4 & 0.374494 \\ 
15 & 0.0001 & 500 & 10 & 20 & 128 & 4 & 4 & 0.397611 \\ 
16 & 0.0001 & 500 & 10 & 24 & 128 & 4 & 4 & 0.302832 \\ 
17 & 0.0001 & 500 & 10 & 32 & 128 & 4 & 4 & 0.330106 \\ 
18 & 0.0001 & 500 & 10 & \textbf{64} & 128 & 4 & 4 & \textbf{0.285905} \\ \hline
19 & 0.0001 & 500 & 10 & 16 & 12  & 4 & 4 & 0.319869 \\ 
20 & 0.0001 & 500 & 10 & 16 & 16  & 4 & 4 & 0.384878 \\ 
21 & 0.0001 & 500 & 10 & 16 & \textbf{20}  & 4 & 4 & \textbf{0.280823} \\ 
22 & 0.0001 & 500 & 10 & 16 & 24  & 4 & 4 & 0.357627 \\ 
23 & 0.0001 & 500 & 10 & 16 & 32  & 4 & 4 & 0.349190 \\ 
24 & 0.0001 & 500 & 10 & 16 & 64  & 4 & 4 & 0.341349 \\ 
25 & 0.0001 & 500 & 10 & 16 & 128 & 4 & 4 & 0.374494 \\ \hline
26 & 0.0001 & 500 & 10 & 16 & 128 & 2 & 4 & 0.315785 \\ 
27 & 0.0001 & 500 & 10 & 16 & 128 & 4 & 4 & 0.374494 \\ 
28 & 0.0001 & 500 & 10 & 16 & 128 & \textbf{6} & 4 & \textbf{0.295493} \\ 
29 & 0.0001 & 500 & 10 & 16 & 128 & 8 & 4 & 0.522469 \\ \hline
30 & 0.0001 & 500 & 10 & 16 & 128 & 4 & 2 & 0.365716 \\ 
31 & 0.0001 & 500 & 10 & 16 & 128 & 4 & 4 & 0.374494 \\ 
32 & 0.0001 & 500 & 10 & 16 & 128 & 4 & 6 & 0.327240 \\ 
33 & 0.0001 & 500 & 10 & 16 & 128 & 4 & \textbf{8} & \textbf{0.282346} \\ \hline
\end{tabular}%
    }
\end{table}

Finally, Table~\ref{tab:hyperparameters} summarizes the best-performing hyperparameter configurations for each model across different problems and training settings. The hyperparameters shown, such as batch size, learning rate, network width, and depth, are the ones that yielded the highest accuracy in our experiments. Each tuple in the table corresponds to the configuration used for the respective temporal approach. This comprehensive overview provides insight into how different models adapt to each problem and highlights the tuning necessary for optimal performance.

\begin{table}[H]
    \centering
    \caption{Best-performing hyperparameter configurations for each model across various PDE problems and temporal approaches. Each entry reports the selected values for different hyperparameters. Values in parentheses correspond to different temporal approaches.}
    \label{tab:hyperparameters}
    \resizebox{\textwidth}{!}{%
    \begin{tabular}{l|ccccccccccccccc}
\multicolumn{1}{c|}{Problem} & \textbf{Method} & \textbf{App.} & \textbf{\(N_{\text{train}}\)} & \textbf{\(N_{\text{test}}\)} & \textbf{\(Bs\)} & \textbf{\(Lr\)} & \textbf{Epoch}   & \textbf{\(M\)} & \textbf{\(W\)} & \textbf{\(W_\mathcal{Q}\)} & \textbf{\(W_\mathcal{H}\)} & \textbf{\(N_l\)} & \textbf{\(N_\mathcal{Q}\)} & \textbf{\(N_\mathcal{H}\)} & \textbf{\(s\)} \\ \hline
\multirow{3}{*}{\begin{tabular}[c]{@{}l@{}}Allen-\\ Cahn\end{tabular}} 
& MHNO   & I (II) (III) & 4050 (3680) (1800) & 100 & 50 & 0.001 & 500  & 12 & 32 & 32  & 8   & 4   & 2  & 2  & 64 \\
& FNO-3d & I (II) (III) & 4050 (3680) (1800) & 100 & 50 & 0.001 & 500  & 8  & 16 & 16  & -  &  4  & 2  & -  & 64 \\
& FNO-2d & I (II) (III) & 4050 (3680) (1800) & 100 & 25 & 0.001 & 500  & 18 & 64 & 20  & -  & 6   & 8  & -  & 64 \\ \hline
\multirow{3}{*}{\begin{tabular}[c]{@{}l@{}}Cahn-\\ Hilliard\end{tabular}}
& MHNO   & I (II) (III) & 6480 (5520) (4200) & 200 & 25 & 0.001 & 500  & 12 & 32 & 64 & 32 & 6  & 2 & 2 & 64 \\
& FNO-3d & I (II) (III) & 6480 (5520) (4200) & 200 & 40 & 0.001 & 500  & 8 & 16 & 16 & -  & 6  & 2  & - & 64 \\
& FNO-2d & I (II) (III) & 6480 (5520) (4200) & 200 & 25 & 0.001 & 500  & 16 & 32 & (128) 32 (32) & - & 6 & 2  & - & 64 \\ \hline
\multirow{3}{*}{\begin{tabular}[c]{@{}l@{}} Swift-\\ Hohenberg\end{tabular}}
& MHNO   & I (II) (III) & 6480 (5520) (4200) & 200 & 25 & 0.001 & 500  & 14 & 32 & 32 & 16 & 4 & 2 & 4 & 64 \\
& FNO-3d & I (II) (III) & 6480 (5520) (4200) & 200 & 40 & 0.001 & 500  & 8 & 16 & 16 & - & 6 & 2  & - & 64 \\
& FNO-2d & I (II) (III) & 6480 (5520) (4200) & 200 & 25 & 0.001 & 500  & 14 & 32 & 32 & - & 4 & 2  & - & 64 \\ \hline
\multirow{3}{*}{\begin{tabular}[c]{@{}l@{}} Phase Field\\ Crystal\end{tabular}}
& MHNO   & I (II) (III) & 6480 (5520) (4200) & 200 & 25 & 0.005 & 500  & 12 & 32 & 32 & 16 & 4 & 2 & 4 & 64 \\
& FNO-3d & I (II) (III) & 6480 (5520) (4200) & 200 & 25 & 0.005 & 500  & 8 & 16 & 16 & - & 4 (6) (6) & 2  & - & 64 \\
& FNO-2d & I (II) (III) & 6480 (5520) (4200) & 200 & 25 & 0.001 & 500  & 16 & 22 (32) (32) & 64  & - & 4 & 2  & - & 64 \\ \hline
\multirow{3}{*}{\begin{tabular}[c]{@{}l@{}} Molecular \\ Beam Epitaxy\end{tabular}}
& MHNO   & I (II) (III) & 6480 (5520) (4200) & 200 & 25 & 0.005 & 500  & 12 & 32 & 32 & 32 & 4 & 2 & 4 & 64 \\
& FNO-3d & I (II) (III) & 6480 (5520) (4200) & 200 & 25 & 0.005 & 500  & 8 & 16 & 16 & - & 4 & 2  & - & 64 \\
& FNO-2d & I (II) (III) & 6480 (5520) (4200) & 200 & 25 & 0.001 & 500  & 14 & 32 & 32 & - & 4 & 2  & - & 64 \\ \hline
\end{tabular}%
    }
\end{table}

\end{appendices}

\printbibliography

\end{document}